\documentclass[sigplan,screen]{acmart}
\setcopyright{acmcopyright} % No copyright notice required for submissions
\acmPrice{15.00}
\acmDOI{10.1145/3519939.3523725}
\acmYear{2022}
\copyrightyear{2022}
\acmSubmissionID{pldi22main-p556-p}
\acmISBN{978-1-4503-9265-5/22/06}
\acmConference[PLDI '22]{Proceedings of the 43rd ACM SIGPLAN International Conference on Programming Language Design and Implementation}{June 13--17, 2022}{San Diego, CA, USA}
\acmBooktitle{Proceedings of the 43rd ACM SIGPLAN International Conference on Programming Language Design and Implementation (PLDI '22), June 13--17, 2022, San Diego, CA, USA}

\usepackage{booktabs}   %% For formal tables:
\usepackage{subcaption} %% For complex figures with subfigures/subcaptions
\usepackage{xcolor}
\usepackage{xspace}
\usepackage{cleveref}
\usepackage{commands}
\usepackage{comment}
\usepackage{hyperref}
\usepackage{graphicx}
\usepackage[utf8]{inputenc}
\usepackage{listings}
\usepackage{balance}
\usepackage{centernot}
\usepackage{xspace}
\usepackage[inference]{semantic}
\usepackage{algorithm}
\usepackage{algpseudocode}
\usepackage{caption}
\usepackage{makecell}
\usepackage{multirow}
\usepackage{stmaryrd}
\usepackage{amsmath}
\usepackage{graphicx}
\usepackage{setspace}
\usepackage{microtype}
\usepackage{mathtools}
\usepackage[english]{babel}

\usepackage{wrapfig}
\usepackage{mathpartir}
\usepackage{paralist}

\usepackage{listings}
\newsavebox\MBox

\ifdefined\withcolor
	 \definecolor{haskellblue}{rgb}{0.0, 0.0, 1.0}
	 \definecolor{haskellstr}{rgb}{0.2, 0.2, 0.6}
	 \definecolor{haskellred}{rgb}{1.0, 0.0, 0.0}
  \definecolor{gray_ulisses}{gray}{0.55}
  \definecolor{castanho_ulisses}{rgb}{0.71,0.33,0.14}
  \definecolor{preto_ulisses}{rgb}{0.41,0.20,0.04}
  \definecolor{green_ulises}{rgb}{0.2,0.75,0}
\else
	\definecolor{haskellblue}{gray}{0.1}
	\definecolor{haskellstr}{gray}{0.1}
	\definecolor{haskellred}{gray}{0.1}
	\definecolor{gray_ulisses}{gray}{0.1}
	\definecolor{castanho_ulisses}{gray}{0.1}
	\definecolor{preto_ulisses}{gray}{0.1}
	\definecolor{green_ulisses}{gray}{0.1}
\fi

\definecolor{lcolor}{gray}{0.0}
\definecolor{lappcolor}{gray}{0.0}
\definecolor{lappascolor}{gray}{0.0}

\def\codesize{\small}

\definecolor{usercolor}{RGB}{47, 46, 51}
\definecolor{liocolor}{RGB}{10, 55, 104}
\definecolor{yesodcolor}{RGB}{11, 104, 51}
\definecolor{dbcolor}{RGB}{47, 46, 51}

\lstdefinelanguage{HaskellUlisses} {
	basicstyle=\ttfamily\small,
	sensitive=true,
	morecomment=[l][\color{gray_ulisses}\sffamily\itshape\codesize]{--},
	morestring=[b]",
	stringstyle=\color{haskellstr},
	basewidth={0.53em},
	showstringspaces=false,
	numberstyle=\codesize,
	numberblanklines=true,
	showspaces=false,
	breaklines=true,
	showtabs=false,
	escapeinside={(*}{*)},
  literate={ {quals}{{$\mathbb{Q}$}}1
             {iquals}{{$\mathbb{Q}$}}2
             {ltsolzero}{{$A_0$}}2
             {band}{{$\textbf{\texttt{and}}$}}2
             {->}{{$\rightarrow$}}1
             {<-}{{$\leftarrow$}}1
             {monotonic}{{monotonic}}9
             {plusminus}{{$\pm$}}2
             {not}{{$\neg\!\!\!$}}2
             {===}{{$\equiv$}}2
             {ς}{{$\varsigma$}}1
             {ε}{{$\epsilon$}}1
             {φ}{{$\phi$}}1
             {canFlowTo}{{$\sqsubseteq$}}1
             {cannotFlowTo}{{$\not\sqsubseteq$}}1
             {meet}{{$\sqcap$}}1
             {jjoin}{{join}}4
             {join}{{$\sqcup$}}1
             {bot}{{$\perp$}}1
             {top}{{$\top$}}1
						 {<=}{{$\leq$}}1
						 {&&}{{$\land$}}1
						 {||}{{$\lor$}}1
             {=>}{{$\Rightarrow$}}1
             {<=>}{{$\Leftrightarrow$}}1
             {GType}{{\tilde{\mathit{\texttt{Type}}}}}4
           },
	emph=
	{[1]
		FilePath,IOError,abs,acos,acosh,all,and,any,appendFile,approxRational,asTypeOf,asin,
		asinh,atan,atan2,atanh,basicIORun,break,catch,ceiling,chr,compare,concat,concatMap,
		const,cos,cosh,curry,cycle,decodeFloat,denominator,digitToInt,div,divMod,drop,
		dropWhile,either,elem,encodeFloat,enumFrom,enumFromThen,enumFromThenTo,enumFromTo,
		error,even,exp,exponent,fail,mapMaybe,filter,flip,floatDigits,floatRadix,floatRange,floor,
		fmap,foldl,foldl1,foldr,foldr1,fromDouble,fromEnum,fromInt,fromInteger,fromIntegral,
		fromRational,fst,gcd,getChar,getContents,getLine,head,id,inRange,index,init,intToDigit,
		interact,ioError,isAlpha,isAlphaNum,isAscii,isControl,isDenormalized,isDigit,isHexDigit,
		isIEEE,isInfinite,isLower,isNaN,isNegativeZero,isOctDigit,isPrint,isSpace,isUpper,iterate,
		last,lcm,length,lex,lexDigits,lexLitChar,lines,log,logBase,lookup,map,mapM,mapM_,max,
		maxBound,posMax,negMax,maximum,maybe,min,minBound,minimum,mod,negate,not,notElem,null,numerator,odd,
		or,ord,pi,pred,primExitWith,print,product,properFraction,putChar,putStr,putStrLn,quot,
		quotRem,range,rangeSize,read,readDec,readFile,readFloat,readHex,readIO,readInt,readList,readLitChar,
		readLn,readOct,readParen,readSigned,reads,readsPrec,realToFrac,recip,rem,repeat,replicate,return,
		round,scaleFloat,scanl,scanl1,scanr,scanr1,seq,sequence,sequence_,show,showChar,showInt,
		showList,showLitChar,showParen,showSigned,showString,shows,showsPrec,significand,signum,sin,
		sinh,snd,span,splitAt,sqrt,subtract,succ,sum,tail,take,takeWhile,tan,tanh,threadToIOResult,toEnum,
		toInt,toInteger,toLower,toRational,toUpper,truncate,uncurry,undefined,unlines,until,unwords,unzip,
		unzip3,userError,words,writeFile,zip,zip3,zipWith,zipWith3,listArray,doParse,empty,for,initTo,
        assert,compose,checkGE,maxEvens,empty,create,get,set,initialize,idVec,fastFib,fibMemo,
        ex1,ex2,ex3,incr,inc,dec,isPos,positives,find,insert,len,size,union,fromList,initUpto,trim,
        insertSort,decsort,qsort,reverse,append,upperCase, ifM, whileM, get, decrM, diff,
        project, select, leq, elts, keys, dkeys, dfun, addKey, pTrue, emptyRD, rFalse,
        	dom, rng, isI, isD, isS, movie1, movie2,  toI, toS, toD, good_titles, runState, ret,
        	update, getCtr, setCtr, ctr, rdCtr, wrCtr, ifTest, whileTest, posCtr, zeroCtr, decr, decCtr,
        	pread , pwrite , plookup , pcontents, pcreateF , pcreateFP, pcreateD, active, caps, pset, eqP,
        	write, contents, alloc, derivP, copyP, createDir, store, copyRec, copySpec,
        	forM_, when, flookup, fread, createDir, pcreateFile, isFile, copyFrame, ?
	},
	emphstyle={[1]\color{haskellblue}},
	emph=
	{[2] Eq, Program, Label, 
	},
	emphstyle={[2]\color{castanho_ulisses}},
	emph=
	{[3]
		case,class,data,deriving,do,else,if,import,in,infixl,infixr,instance,let,
		module,of,primitive,then,refinement,type,where,forall,bound,and,
		measure,reflect,predicate, assume, return
	},
	emphstyle={[3]\color{preto_ulisses}\textbf},
	emph=
	{[4]
		quot,rem,div,mod,elem,notElem,seq
	},
	emphstyle={[4]\color{castanho_ulisses}\textbf},
	emph=
	{[5]
		EQ,GT,LT,Left,Right
	},
	emphstyle={[5]\color{preto_ulisses}\textbf},
	emph=
	{[6]
	    axiomatize, measure, inline, return
	},
	emphstyle={[6]\color{lcolor}}
}

\lstnewenvironment{code}
{\lstset{language=HaskellUlisses}}
{}

\lstnewenvironment{mcode}
{\lstset{language=HaskellUlisses,columns=fullflexible,keepspaces,mathescape,stepnumber=1,firstnumber=1}}% ,numberfirstline=true,numberstyle={\scriptsize\it},numbers=left}}
{}

\lstMakeShortInline[language=HaskellUlisses,mathescape,keepspaces,mathescape,basicstyle=\ttfamily\small,breakatwhitespace]@

\lstdefinelanguage{Pseudo} {
	basicstyle=\ttfamily\codesize,
	sensitive=true,
  mathescape=true,
	morecomment=[l][\color{gray_ulisses}\ttfamily\codesize]{--},
	morecomment=[s][\color{gray_ulisses}\ttfamily\codesize]{\{-}{-\}},
	morestring=[b]",
	showstringspaces=false,
	numberstyle=\codesize,
	numberblanklines=true,
	showspaces=false,
	breaklines=true,
	showtabs=false
}

\algrenewcommand\algorithmicindent{0.8em}

\definecolor{ForestGreen}{RGB}{34,139,34}
\definecolor{Airforceblue}{rgb}{0.36, 0.54, 0.66}
\definecolor{Burgundy}{rgb}{0.5, 0.0, 0.13}

\begin{document}

\author{Sankha Narayan Guria}
\affiliation{
    \institution{University of Maryland}
    \city{College Park}
    \country{USA}
}
\author{Niki Vazou}
\affiliation{
    \institution{IMDEA Software Institute}
    \city{Madrid}
    \country{Spain}
}
\author{Marco Guarnieri}
\affiliation{
    \institution{IMDEA Software Institute}
    \city{Madrid}
    \country{Spain}
}
\author{James Parker}
\affiliation{
    \institution{Galois, Inc.}
    \city{Arlington}
    \country{USA}
}

\title[\tool: Approximated Knowledge Synthesis with Refinement Types for Declassification]{\tool: Approximated Knowledge Synthesis with Refinement Types for Declassification}


\begin{abstract}
Non-interference is a popular way to enforce confidentiality of
sensitive data. 
However, declassification of sensitive information is often needed in realistic applications but breaks non-interference.
We present \tool, an 
\underline{a}pproximate k\underline{no}wledge \underline{sy}nthesizer
for quantitative declassification policies. \tool uses refinement types
to automatically construct machine checked over- and
under-approximations of attacker knowledge for boolean queries on
multi-integer secrets.
It also provides an \texttt{AnosyT} monad to track the attacker knowledge over
multiple declassification queries and checks for violations against
user-specified policies in information
flow control applications.
We implement a prototype of \tool and show that it is precise 
and permissive: up to 14 declassification queries are permitted 
before a policy violation occurs using the powerset of intervals domain.
\end{abstract}

\maketitle

\section{Introduction}
\label{sec:intro}

%%% NI 
Information flow control (IFC)~\cite{SabelfeldM03} systems protect the confidentiality of sensitive data during program execution.
They do so by enforcing a property called \emph{non-interference} which ensures the absence of leaks of secret information (say, a user location) through public observations (say, information being sent to the network socket).
%

%% NI is restrictive --> Declassification is needed
Real-world programs, however, often need to reveal information about sensitive data.
For instance, a location based web application needs to suggest restaurants 
or friends that are  nearby the \texttt{Secret} user location. 
Such computations, which leak information about the \texttt{Secret} location, would be prevented by IFC systems that enforce non-interference. % checker violates non-interference since the \texttt{Public} user learns information about the \texttt{Secret} password, 
To support them, IFC systems provide \emph{declassification} statements~\cite{sabelfeld2009declassification} that can be used to weaken non-interference by allowing the selective disclosure of some \texttt{Secret} information.\looseness=-1

Declassification statements, however, are typically part of an application's trusted computing base and developers are responsible for properly declassifying information.
In particular, mistakes in declassification statements can easily compromise a system's security because declassified information bypasses standard IFC checks.
Implementing declassification statements can be difficult for developers to implement correctly.
For example, \citet{gonzalez2021unique} showed that non-Personally Identifiable Information (PII) in an advertising system could be combined to uniquely identify and target an individual.
Developers may declassify seemingly non-sensitive non-PII, but accidentally leak sensitive information about a person's identity.
% 
% private user information can be leaked with unsafe
% declassification when nano-targeting users with non-personally
% identifiable information in an advertising
% system~\cite{gonzalez2021unique}.
% \mg{Do we want to say that one might declassify seemingly non-sensitive non-PIIs and still leak sensitive stuff?}
%
Instead of trusting the developer to correctly declassify information, an alternative approach is to enforce \emph{declassification policies}~\cite{chong2004security} that regulate the use of declassification statements. % information can be declassified (and under which conditions).

In this paper, we present \tool{}, % an \underline{a}bstract k\underline{no}wledge \underline{sy}nthesis 
a framework for enforcing \emph{declassification policies} on IFC systems where policies regulate \emph{what} information can be declassified~\cite{sabelfeld2009declassification} by limiting the amount of information an attacker could learn from the declassification statements.
Specifically, declassification policies are expressed as constraints over \emph{knowledge}~\cite{askarov2007gradual}, which semantically characterizes the set of secrets an attacker considers possible given the prior declassification statements.
To enforce such policies, we develop (1) a novel encoding of knowledge approximations using Liquid Haskell's~\cite{liquidHaskell} refinement types which we use to (2) automatically synthesize correct-by-construction knowledge approximations for Haskell queries. 
We then (3) implement and (4) evaluate a 
knowledge tracking and policy enforcing declassification function 
that can easily extend existing IFC monadic systems. 
Next, we discuss these four contributions in detail.

\paragraph{Verified knowledge approximations}
We define a novel encoding for knowledge approximations over abstract domains using Liquid Haskell (\S~\ref{sec:encoding}). 
The novelty of our encoding is that approximation data types 
are indexed by two predicates that respectively capture the properties of 
elements inside and outside of the domain. 
Using these indexes, we encode correctness 
of over- and under-approximations, without using quantification, permitting SMT-decidable verification.
With this encoding, we implement and machine check Haskell 
approximations of two abstract domains: 
intervals over multi-dimensional spaces (where each dimension is abstracted using an interval)
and powersets on these intervals, that increase the precision of our approximations. 
This verified knowledge encoding  is general and can be used, beyond declassification, also as building block for dynamic~\cite{guarnieri2019information, van2015very}, probabilistic~\cite{SweetTS0M18,mardziel-beliefpol-2013,guarnieri2017securing,kucera2017synthesis}, and quantitative policies~\cite{backesleak09,KopfR10}.

\paragraph{Synthesis of knowledge approximations}
We develop a novel approach for automatically synthesizing correct-by-construction posteriors given any prior knowledge and user-specified boolean query over multi-dimensional integer secret values (\S~\ref{sec:synthesis}).
Our approach combines type-based sketching with SMT-based synthesis and it is implemented as a Haskell compiler plugin, \ie it operates at compile-time on Haskell programs. 
Given a user-defined query, \tool
generates a synthesis template (a so-called sketch)  where the values of the abstract domain elements are left as \emph{holes} to be filled later with values, combined with the correctness specification encoded as refinement types.
It then reduces the high-level correctness property into integer constraints on bounds of the abstract domain elements and uses an SMT solver to synthesize \emph{optimal} correct-by-construction values. 
Replacing these values in the sketch, \tool synthesizes Haskell executable programs of the approximated  knowledge and 
automatically checks their correctness
with Liquid Haskell.\looseness=-1 % of the synthesized programs. 

\paragraph{Enforcing declassification policies}
We implement a policy-based declassification function 
that can be used by any monadic Haskell IFC framework (\S~\ref{sec:overview}, \S~\ref{sec:monad}).
In this setting, users write declassification policies as Haskell functions that constrain the (approximated) attacker knowledge, whereas declassification queries are written as regular Haskell functions over secret data. 
At compile time, \tool synthesizes and verifies the knowledge approximations for all declassification queries. 
At runtime, declassification is called in the  @AnosyT@ monad that  
tracks knowledge over multiple declassification queries and 
checks, using the synthesized knowledge approximations, whether performing the declassification would lead to violating the user-specified policy.
Importantly, @AnosyT@ is defined as a monad transformer, thus can be staged on top of 
 existing IFC monads like LIO~\cite{stefan2011flexible} and STORM~\cite{strorm}.

\paragraph{Evaluation}
We evaluated precision and running time of \tool using two benchmarks (\S~\ref{sec:evaluation}). First, we compared with \texttt{Prob}'s~\cite{mardziel-beliefpol-2013} 
benchmark suite to conclude that \tool is slower but more precise. 
Second, to demonstrate \tool enables secure declassification of sequential queries, we evaluate how many queries \tool allows to declassify before a policy violation. For the interval abstract domain, we found a policy violation was detected after a maximum of 7 queries and after 14 queries for the more precise powerset domain.

\section{Overview}
\label{sec:overview}
We start by motivating the need for declassification policies (\S~\ref{subsec:downgrade}): 
repeated downgrades can weaken non-interference until leaking the secret is allowed. 
Next, we present how the knowledge revealed by queries 
can be computed (\S~\ref{subsec:knowledge}). 
Finally (\S~\ref{subsec:overview:anosy}), 
we describe how \tool synthesizes correct-by-construction 
knowledge, by combining
refinement types, SMT-based synthesis, and metaprogramming. 

\subsection{Motivation: Bounded Downgrades}
\label{subsec:downgrade}
\paragraph{Secure Monads}
IFC systems, \eg LIO~\cite{stefan2011flexible} and LWeb~\cite{parker2019lweb}, 
define a \textit{secure} monad to ensure that 
security policies are enforced over sensitive data, 
like a user's physical location. 
For instance here, we define the data type @UserLoc@ to capture 
the user location as its @x@ and @y@ coordinates. 
\begin{mcode}
  data UserLoc = UserLoc {x :: Int, y :: Int}
\end{mcode}
A \texttt{Secure} monad will return such a location wrapped in a 
protected ``box'' to ensure that only 
code with sufficient privileges
can inspect it. 
For example, a function that gets the user's location 
will return a protected value: 
\begin{mcode}
  getUserLoc :: User -> Secure (Protected UserLoc)
\end{mcode}
In the @LIO@ monad, for example, data are protected by a security label data type
and the monad ensures, based on the application, that 
only the intended agents can observe (or unlabel) 
the user's exact location. 

\paragraph{Queries}
%
% In the following, 
% We define a query to be any ...
A \textit{query} is any 
boolean function over \textit{secret} values. 
As an example, we consider the user location to be the secret 
value and the @nearby@ function below checks proximity to 
this secret value from @(x_org, y_org)@.
\begin{mcode}
  type $\cdom$ = UserLoc

  nearby :: (Int, Int) -> $\cdom$ -> Bool
  nearby (x_org, y_org) (UserLoc x y) 
    = abs (x - x_org) + abs (y - y_org) <= 100 
    where abs i = if i < 0 then -i else i
\end{mcode}
The @nearby@ query is using Manhattan distance to check if 
a user is located within @100@ units of the input origin location. 

\paragraph{Downgrades}
Even though locations protected by the \texttt{Secure} monad cannot be inspected
by unprivileged code, in practice many applications need to allow selective leaks of secret information to unprivileged code.
For instance, many web  applications 
need to check location of users to provide useful information,  
such as restaurant, friend, or dating suggestions 
that are physically \texttt{nearby} the user.

The \texttt{showAdNear} function  below shows a restaurant 
advertisement to the user only if they are nearby. 
To do so, the function uses \texttt{downgrade} (from the \texttt{Secure} monad) to downgrade (to \texttt{public}) the result of the \texttt{nearby} check over the protected user location. 
\begin{mcode}
  downgrade  :: Protected $\cdom$ -> ($\cdom$ -> Bool) 
             -> Secure Bool 
 
  showAd     :: User -> Restaurant -> Secure () 
  showAdNear :: User -> Restaurant -> Secure () 
  showAdNear user res = do 
    ul     <- getUserLoc user 
    isNear <- downgrade ul (nearby (res_loc res))
    if isNear then showAd user res else return ()
\end{mcode}
Downgrades are a common feature of real-world IFC systems. 
For example, in LIO downgrades happen with the
@unlabelTCB@ trusted codebase function, which is exposed to
the application developers. 
At the same time, downgraded information bypasses security checks by design.
In the code above, \texttt{isNear} is unprotected and can now be leaked to an attacker.
Therefore, declassification statements need to be correctly placed to avoid unintended leaks of information that would bypass IFC enforcement.

\paragraph{Declassification knowledge}
\begin{figure*}
\centering
\begin{subfigure}[b]{0.33\textwidth}
  \centering
\includegraphics[width=0.7\textwidth]{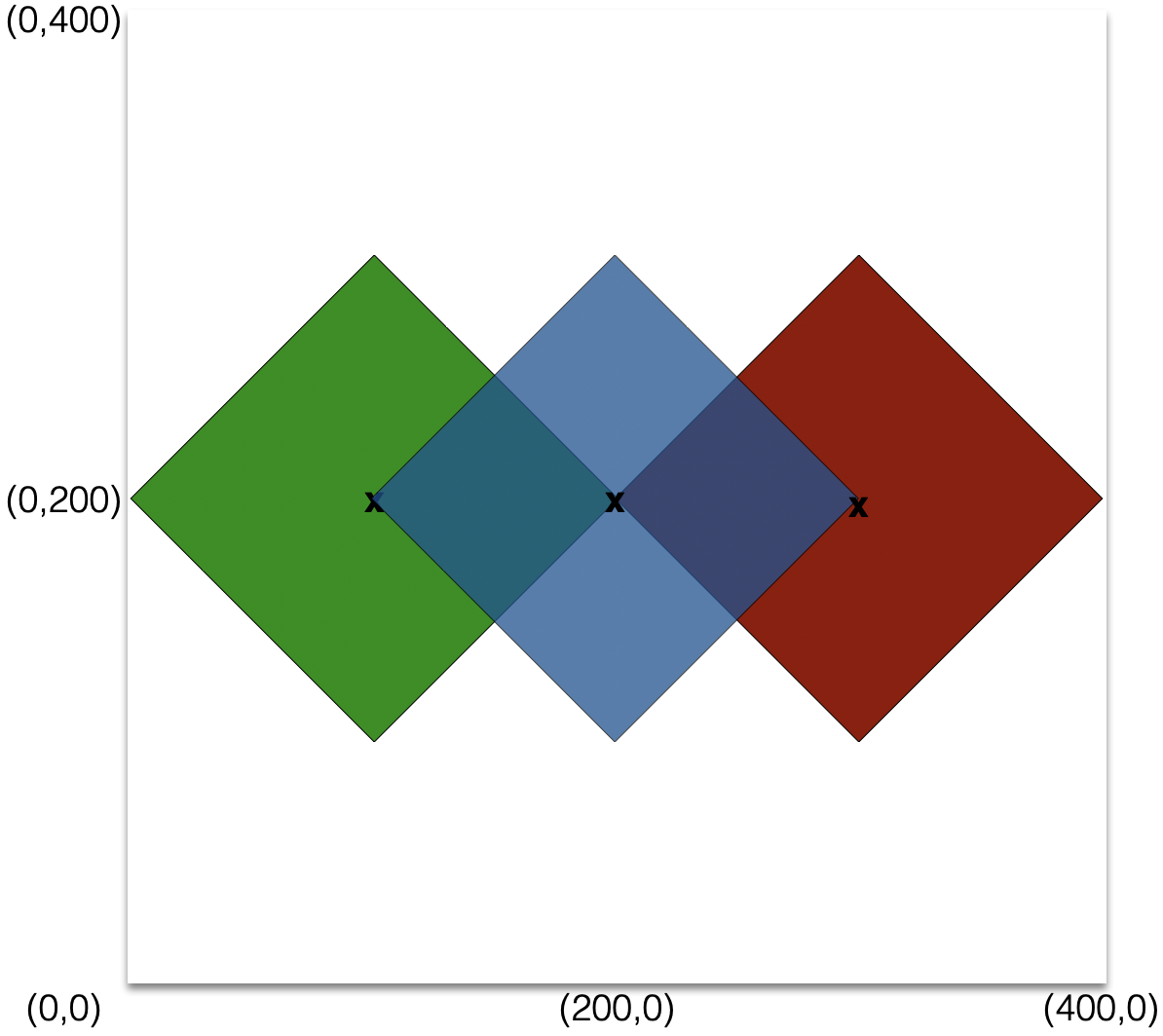}
\caption{Posteriors on x = 
  \textcolor{ForestGreen}{200}, \textcolor{Airforceblue}{300}, 
   \textcolor{Burgundy}{400} and y = 200.}
\label{fig:knowledge}
\end{subfigure}
\begin{subfigure}[b]{0.33\textwidth}
  \centering
  \includegraphics[width=0.7\textwidth]{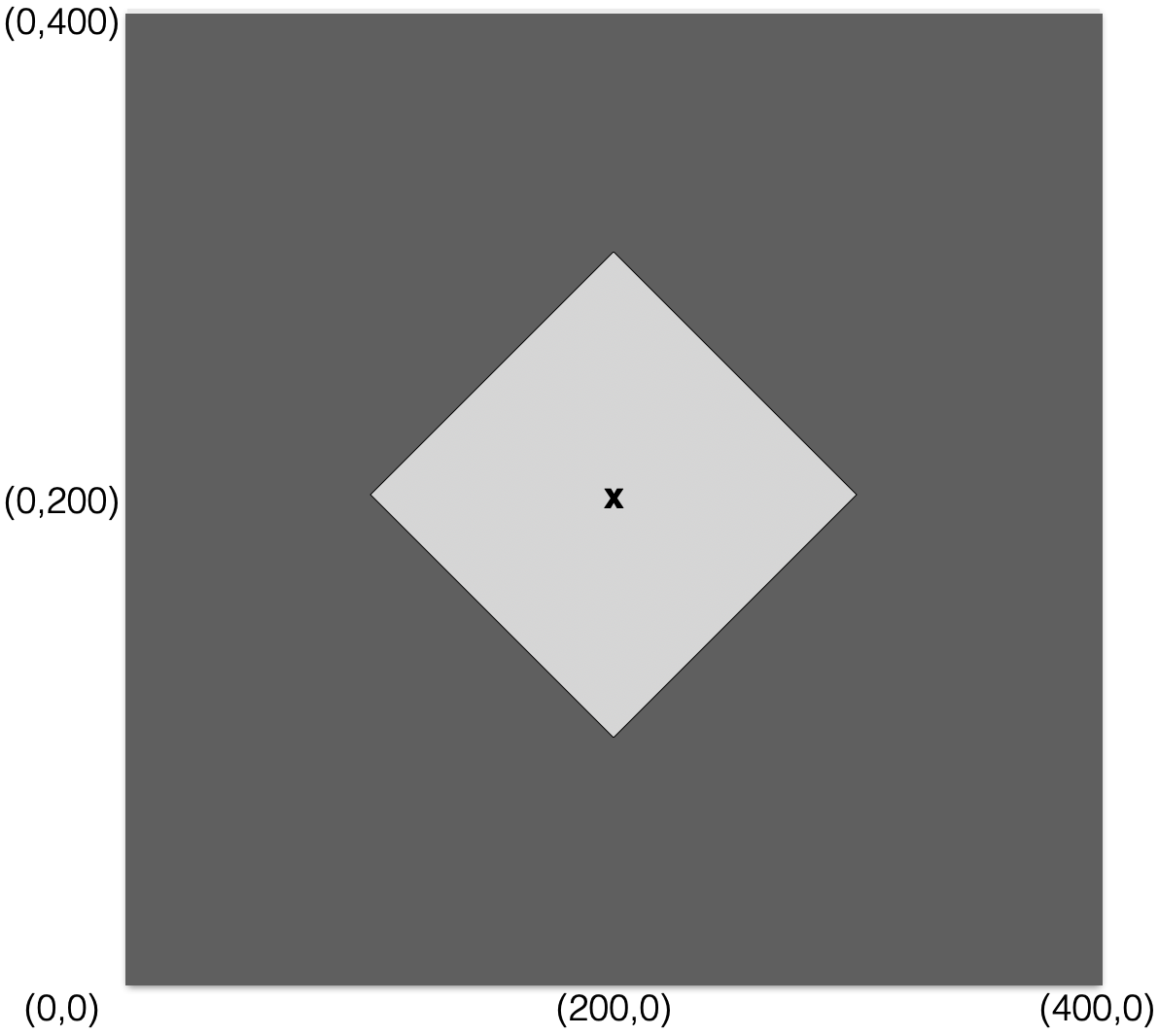}
\caption{Indistinguishability Sets}
\label{fig:nearby:real}
\end{subfigure}%
\begin{subfigure}[b]{0.33\textwidth}
  \centering
\includegraphics[width=0.7\textwidth]{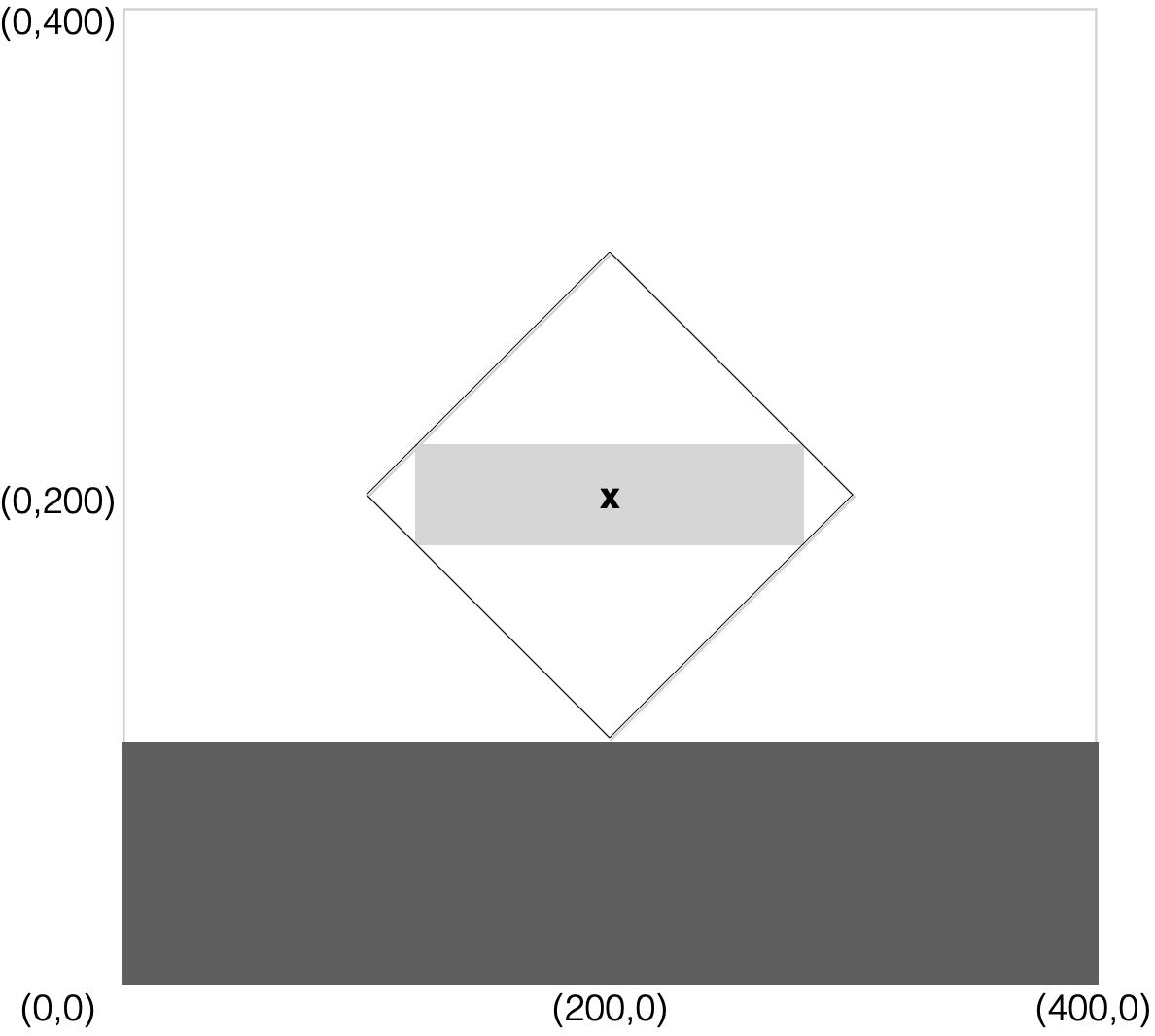}
\caption{Under-Approximation}
\label{fig:nearby:under}
\end{subfigure}%
\caption{Posteriors, Indistinguishability Sets and their
Approximations with respect to \texttt{nearby} query.
}
\label{fig:nearby-example}
\end{figure*}
To semantically characterize the information declassified by \texttt{downgrade}s, we use the notion of \emph{attacker knowledge}~\cite{askarov2007gradual}, \ie the set of secrets that are consistent with an attacker's observations, where attackers can observe the results of \texttt{downgrade}. 
That is, we consider the worst-case scenario where any declassified information is \emph{always} leaked to an attacker.
This knowledge can be refined by consecutively running downgrade queries and ultimately can reveal the exact value of the secret. 
In the example below, a piece of code downgrades two queries asking 
if the user is located nearby to both 
@(200,200)@ and @(400,200)@ to infer if the 
exact user location is @(300,200)@.
\begin{mcode}
  secret <- getUserLoc user
  kn1    <- downgrade secret (nearby (200,200))
  kn2    <- downgrade secret (nearby (400,200))
  -- if kn1 && kn2, then secret = (300,200)
\end{mcode}
% %
The \textit{posterior} is the knowledge obtained after executing a query. 
Consider again the code above. 
If @nearby (200,200)@  is true, the knowledge after the first downgrade statement is the green region of Figure~\ref{fig:knowledge}. 
Using this information as \textit{prior} knowledge for the second downgrade query, which asks @nearby (400,200)@, might result in a knowledge containing only the user location @(300,200)@, 
\ie the intersection of the green and red 
posterior knowledge regions.

\paragraph{Quantitative Policies}
A \textit{quantitative policy}
is a predicate on knowledge which, for instance, ensures that 
the accumulated knowledge is not specific enough,
\ie the secret cannot be revealed. 
As an example, @qpolicy@ below states that 
the knowledge should contain at least @100@ values. 
%
%   qpolicy :: $\adom$ -> Bool
\begin{mcode}
   qpolicy dom = size dom > 100
\end{mcode}

This policy will allow declassifying @nearby (200,200)@ and @nearby (300,200)@,
since the intersections of the green and blue regions in
Figure~\ref{fig:knowledge} contain at least 100 potential locations, but not
@nearby (400,200)@ since the resulting knowledge contains exactly one secret.

\paragraph{Bounded Downgrade}
We define a bounded downgrade operator that allows 
the computation of queries on secret data, 
while enforcing quantitative policies. 
For example, the operator tracks declassification knowledge during an execution and allows downgrading the @nearby (200,200)@
and @nearby (300,200)@ queries, but  terminates with an
 error on the sequence of @nearby (200,200)@
and @nearby (400,200)@. 

The downgrade operation is the method of the 
@AnosyT@ monad (\S~\ref{sec:monad})
which is defined as a state monad transformer. 
As a state monad, it preserves the protected secret, 
the quantitative policy, and 
the prior declassification knowledge. % that an 
% attacker can have accumulated.?
To @downgrade@ a new query, the monad checks 
if the posterior knowledge of this 
query satisfies the policy. 
If not, it terminates with a policy violation error. 
Otherwise, it updates the knowledge to the posterior 
and returns the query result.  
Since @AnosyT@ is also a monad transformer, it can be combined with existing security monads, which provide the underlying IFC enforcement mechanism, to enrich them with extra quantitative guarantees on the inevitable downgrades.

\subsection{Approximating knowledge from queries}
\label{subsec:knowledge}

Precisely computing, representing, and checking quantitative policies 
over a (potentially infinite) 
knowledge requires reasoning about all points in the input space, which is an
uncomputable task in general. 
So, we use abstract domains (here
intervals~\cite{cousot1976static}) to approximate knowledge. 

\paragraph{Indistinguishability sets}

The proximity query @nearby (200,200)@  partitions the space of secret 
locations into
two partitions (for the two possible responses: @True@ and @False@), called
\textit{indistinguishability sets} (\indsets), 
\ie all secrets in each partition produce the same result for the
 query. Figure~\ref{fig:nearby:real} depicts the two \indsets 
for our query. The inner diamond---depicted in light
gray---is the \indset for the result @True@, \ie all its elements
respond @True@ to the query. In contrast, the
outer region---depicted in dark gray---is the \indset for
@False@. Figure~\ref{fig:nearby:under} depicts 
the under-approximated (\ie subset) \indsets for the query as defined by 
% the @under_indset@ zero argument function below: 
@under_indset@: 
\begin{mcode}
  data $\adom$Int = $\adom$Int {lower :: Int, upper :: Int}
  data $\adom$ = $\adom$ [$\adom$Int]

  under_indset :: ($\adom$, $\adom$)
  under_indset = ($\adom$ [$\adom$Int 121 279, $\adom$Int 179 221],
                  $\adom$ [$\adom$Int   0 400, $\adom$Int   0  99])
\end{mcode}
The data @$\adom$Int@ abstracts integers as intervals between 
a lower and an upper value. 
$\adom$ is our abstract knowledge data type that is defined as a list of
abstract integers, which can be used to abstract data with any number of integer
fields. 
The @under_indset@ is a tuple, where the first element corresponds to the
@True@ response and the second element to the @False@ response. It says
all secrets in $x \in [121, 279]$ and $y \in [179, 221]$ evaluate to
@True@ for the query and all secrets in $x \in [0, 400]$ and $y \in [0,
99]$ evaluate to @False@.

\paragraph{Knowledge under-approximation.}
We use \indsets to compute the \emph{posterior} knowledge after the query, \ie
the set of secrets considered possible after observing the query result.
To do so,   
we simply take the intersection $\cap$ of the prior knowledge with the
\indsets associated with the query~\cite{askarov2007gradual,
backesleak09}. If the intersection happens with the exact \indsets, then
we derive the exact posterior. For our example, we intersect with the
under-approximate \indset to produce an under-approximation of the
posterior knowledge \ie an under-approximation of the information learned  when observing the query result. 
\begin{mcode}
  underapprox :: $\adom$ ->  ($\adom$, $\adom$)
  underapprox p = (p $\cap$ trueInd, p $\cap$ falseInd)
  where (trueInd, falseInd) = under_indset
\end{mcode}
The intersection $\cap$ refers 
to the set-theoretic intersection of two domains. We formally define
these operations in \S~\ref{sec:encoding}.

\subsection{Verification and Correct-by-Construction Synthesis of Knowledge}
\label{subsec:overview:anosy}
Our goal is to generate a knowledge approximation 
for each downgrade query, which 
as shown by our @nearby@ example is a strenuous and error prone 
process. 
To automate this process \tool uses
refinement types, metaprogramming, and SMT-based synthesis to 
automatically generate correct-by-construction knowledge approximations of queries in four steps. 
First, for each query \tool generates 
a refinement type specification that denotes knowledge approximation. 
Next, it uses metaprogramming to generate
a partial program, called a sketch, \ie a function definition with holes (to be eventually substituted with terms) that computes the knowledge.
Then, it uses an SMT solver to fill in the
integer value holes in the sketch. 
Finally, it delegates to Liquid Haskell's refinement type checker to verify that our synthesized 
knowledge indeed satisfies its specification. 

Here, we explain a simplified version of these steps for our @nearby (200,200)@ example query. 

\paragraph{Step I: Refinement Type Specifications}
\tool uses abstract refinement types to index 
abstract domains with a predicate that all its elements should 
satisfy (\S~\ref{sec:encoding}). 
For example, @$\adom$ <{\l -> 0 < l}>@
denotes the abstract domain whose elements are positive values. 
Using this abstraction, \tool specifies the 
\indset and knowledge approximations:

\begin{mcode}
  under_indset :: ($\adom$ <{\l ->   nearby l}>, 
                   $\adom$ <{\l -> $\lnot$ nearby l}>)

  underapprox :: p: $\adom$ 
              -> ($\adom$ <{\x ->   nearby x && (x $\in$ p)}>,
                  $\adom$ <{\x -> $\lnot$ nearby x && (x $\in$ p)}>)
\end{mcode}
@under_indset@ returns a tuple of abstract domains. 
The first abstract domain can only contain elements that satisfy 
the query and the second that falsify it. 
The function @underapprox@ computes the posterior given some
prior knowledge @p@. The posterior is further refined to contain only
elements that originally existed in the prior knowledge. 

\paragraph{Step II: Sketch Generation}
Using syntax directed meta-programming \tool 
defines @underapprox@ as in \S~\ref{subsec:knowledge} 
to be the intersection of the \indset and the prior knowledge. 
For the definition of the \indset it relies on the secret type 
to be translated to generate a sketch with integer value holes. 
Since @UserLoc@ contains 
two integer fields, the sketch~\cite{solar2006combinatorial} for
@under_indset@ is the following, where all @l@ and @u@ are holes:
\begin{mcode}
  under_indset = ($\adom$ [$\adom$Int l$_{t1}$ u$_{t1}$, $\adom$Int l$_{t2}$ u$_{t2}$],
                  $\adom$ [$\adom$Int l$_{f1}$ u$_{f1}$, $\adom$Int l$_{f2}$ u$_{f2}$])
\end{mcode}

\paragraph{Step III: SMT-Based Synthesis}
Finally, it combines the refinement type with the program sketch 
to generate, using an SMT solver, solutions for the integer holes (\S~\ref{sec:synthesis}).
By combining values from the above sketch for @under_indset@ with its
refinement type, the below constraints are generated: 
\begin{align}
\forall x, y.\ & l_{t1} \leq x \leq u_{t1}\ \land\ l_{t2} \leq y \leq u_{t2} \implies nearby(x, y) \tag{Under-approx, \textsf{True}}\\
\forall x, y.\ & l_{f1} \leq x \leq u_{f1}\ \land\ l_{f2} \leq y \leq u_{f2} \implies \lnot nearby(x, y) \tag{Under-approx, \textsf{False}}\nonumber
\end{align}

The first constraint indicates all points in the domain should satisfy
the @nearby@ function, whereas the second constraint means the all points
inside the domain should not satisfy the nearby function. The definition
of @nearby(200,200)@ and the @abs@ function is
mechanically translated to logic as follows:
\begin{align*}
query(x, y) &= abs(x - 200) + abs(y - 200) \leq 100\\
abs(i) &= \texttt{if}\ i < 0\ \texttt{then}\ -i\ \texttt{else}\ i  
\end{align*}

These constraints have multiple correct solutions, but, for precision,
\tool prefers the tightest bounds.
Specifically, when under-approximating, it aims for the maximal 
domain that satisfies the above two constraints. 
%This translates to
%generating the maximal, most precise domain . 
%, and minimal domain when over-approximating. 
\tool uses Z3~\cite{bjorner2015nuz} as the SMT
solver of choice because it supports optimization directives for
maximizing $u_1 - l_1$ and $u_2 - l_2$ together, for both the true and false cases.
Finally, it uses the SMT synthesized solutions to fill in the holes 
and derive complete programs, like the @under_indset@ of~\S~\ref{subsec:knowledge}.

\paragraph{Step IV: Knowledge Verification}
\tool uses LiquidHaskell to verify the synthesized result. 
To achieve this step, we implemented (\S~\ref{sec:encoding}) verified
abstract domains for intervals and their powersets that, as shown in our
evaluation~\S~\ref{sec:evaluation}, greatly increase the precision of
the abstractions.
These implementations are independent of the synthesis step and can %directly 
be used to verify manually user-written, knowledge approximations as well.

\section{Bounded Downgrade}
\label{sec:monad}

Here we present the bounded downgrade operation, 
first by an example that showcases how 
downgrades that violate the quantitative declassification policy are rejected,
next by providing its exact implementation, 
and finally by showing correctness of policy enforcement. 

\paragraph{Bounded Downgrade by Example}
The bounded downgrade function checks, before running a downgrade query using the underlying @Secure@ monad, that the approximation of the revealed knowledge satisfies the quantitative policy. 
To do so, it preserves a state that 
maps each secret that has been involved in downgrading operations to 
its current knowledge. 
As an example, below we present 
how the knowledge is updated to prevent 
the example from~\S~\ref{subsec:downgrade}.
\begin{mcode}
  secret <- lift (getUserLoc user)
  -- secret  = Protected (UserLoc 300 200)
  -- secrets = []
  r1 <- downgrade secret "nearby (200,200)"
  -- secrets = [(secret, post1 = {121...279,179...221})], |post1| = 6837
  r2 <- downgrade secret "nearby (300,200)"
  -- secrets = [(secret, post2 = {221...279,179...221})], |post2| = 2537
  r3 <- downgrade secret "nearby (400,200)"
  -- secrets = [(secret, post3 = {$\emptyset$         , 179 ... 221})], |post3| = 0
  -- Policy Violation Error  
\end{mcode}
The user location is taken by lifting the @getUserLoc@ 
function of the underlying monad (any computation of the underlying monad can be lifted).
Assume that the user is located at @(300,200)@. 
Originally, there is no prior knowledge for this secret (and protected)
location, \ie the @secrets@ map associating secrets to knowledge approximations is empty. 
After downgrading the @nearby (200,200)@ query 
(which as we will explain next, is passed to @downgrade@ as a string) we get the posterior @post1@
with size @6837@. Since this size is greater than @100@, the @qpolicy@
(defined in~\S~\ref{subsec:downgrade}) is satisfied and the result of the query (here true) is returned by 
the bounded downgrade. 
Similarly, downgrade of the @nearby (300,200)@ query refines the posterior 
to size @2537@. 
But, when downgrading the @nearby (400,200)@ query the posterior size becomes zero, 
thus our system will refuse to perform the query (and downgrading its result) and return a policy violation error, 
instead of risking the leak of the secret.

\paragraph{Definition of Bounded Downgrade}
\begin{figure}
\begin{mcode}
  type AnosyT a s m = StateT (AState a s) m 

  data AState a s = AState {
    policy   :: a -> Bool,
    secrets  :: Map s a,
    queries  :: Map String (QInfo a s)}

  data QInfo a s = QInfo { 
    query  :: s -> Bool,
    approx :: p: a 
           -> (a <{\x ->   query x && (x $\in$ p)}>,
               a <{\x -> $\lnot$ query x && (x $\in$ p)}>)}

  class Unprotectable p where 
    unprotect :: p t -> t

  downgrade :: ( Monad m, Unprotectable protected
               , AbstractDomain a s) -- Defined in $\S$ 4.1
            => protected s 
            -> String -- (s -> Bool)
            -> AnosyT a s m Bool
  downgrade secret' qName = do
    st <- get
    let qinfo  = lookup qName (queries st)
    if isJust qinfo then do
      let secret = unprotect secret' 
      let prior  = fromMaybe $\top$ 
                 (*\$*) lookup secret (secrets st)
      let (QInfo query approx) = fromJust qinfo
      let (postT, postF) = approx prior
      if policy st postT && policy st postF then do
        let response = query secret
        let posterior = if response then postT 
                        else postF
        modify (*\$*) \st -> st {secrets = 
          insert secret posterior (secrets st)}
        return (*\$*) response
      else throwError "Policy Violation"
    else throwError ("Can't downgrade " ++ qName)
\end{mcode}
\caption{Implementation of bounded \texttt{downgrade}.}
\label{fig:monad-downgrade}
\end{figure}
Figure~\ref{fig:monad-downgrade} presents the definition of 
the bounded @downgrade@ function. 
It takes as input a protected secret, 
which should be able to get @unprotected@ by an instance of 
the @Unprotectable@ class, 
a string that uniquely determines the query to be executed, and 
returns a boolean value in the @AnosyT@ state monad transformer~\cite{10.1145/199448.199528}. 
As discussed in~\S~\ref{subsec:downgrade}, we used a transformer to  
stage our downgrade on top of an existing secure monad. 

The state of Anosy @AState@ contains 
the quantitative policy, 
the map @secrets@ of secret values to their current knowledge,
and the map @queries@ that maps strings that represent queries 
to query information @QInfo@ that, in turn, contain both the query itself 
and an under-approximation function (like the synthesized @underapprox@$\!$) 
that given the prior knowledge approximates the posterior, 
after the query is executed. 
Even though tracking of multiple secrets is permitted, 
we require 
all the secrets and abstractions to have the same type; this limitation can be lifted using heterogeneous 
collections~\cite{10.1145/1017472.1017488}.\looseness=-1

Having access to this state, 
@downgrade@ will throw an error if 
it cannot find the query information of the string input, 
since it has no way to generate the posterior knowledge\footnote{On-the-fly synthesis albeit possible would be very expensive.}.
% (Note that on-the-fly synthesis is possible but would be very expensive.)
Then, it will compute the posterior and throw an error 
if it violates the quantitative policy. 
Otherwise, it will update the posterior of the secret 
and return the result of the query.
Note that detection of violations of the quantitative policy is independent of the actual secret value.

% Alternatively, one can use a
% termination-insensitive monad with LiquidHaskell's termination checker
% to ensure the absence of termination leaks.

\paragraph{Correctness: Policy Enforcement}
Suppose a secret @s@ that has been downgraded
$n$ times by the queries @query$_1$@, \ldots, @query$_n$@. 
After each downgrade, the knowledge is refined.
So, starting from the top knowledge 
($\mathcal{K}_0 \doteq \top$), 
 after $n$ queries, the
 knowledge evolves as follows:
$\mathcal{K}_0\subseteq\mathcal{K}_1\subseteq \dots 
\subseteq\mathcal{K}_i\subseteq \dots \subseteq\mathcal{K}_n$, 
where $\mathcal{K}_i = \mathcal{K}_{i-1} 
\cap \{ x \mid \text{query}_i\ x =  \text{query}_i\ \texttt{s} \}$.

We can show that for each $i$-th downgrade of the secret @s@,  
there exists a posterior $\mathcal{P}_i$ so that
@(s,$\mathcal{P}_i$)@ is in the @secrets@ map and also 
$\mathcal{P}_i$
is an under-approximation of the knowledge  $\mathcal{K}_i$, 
that is $\mathcal{P}_i \subseteq \mathcal{K}_i$.
The proof goes by induction on $i$, 
assuming that the attacker and the @downgrade@ implementation 
start from the same $\top$ knowledge, 
and the inductive step relies on the specification of the 
@approx@ function and the way @downgrades@ modifies @secrets@, 
\ie using @postT@ or @postF@ depending on the response of the query.

Thus if our quantitative policy 
enforces a lower bound on the size of the leaked knowledge, 
(\eg @qpolicy dom = size dom > k@) it is correctly enforced 
by @downgrade@: 
since  $\mathcal{P}_i \subseteq \mathcal{K}_i$, then
@qpolicy $\mathcal{P}_i$@ implies 
@qpolicy $\mathcal{K}_i$@ at each stage of the execution. 
Note that for correctness of policy enforcement, 
the policy should be an increasing function in the size of 
the input for underapproximations. The exact definition of such 
a policy domain specific language is left as a future work. 
Further, even though our implementation 
can trace knowledge overapproximations, 
we have not yet studied applications or policy enforcement 
for this case.  
Last but not least, it is important that the policy is checked 
irrespective of the query result, \ie on both @postT@ and @postF@, 
to prevent potential leaks due to the security decision.

\paragraph{Security Guarantees}

\tool enforces declassification policies that limit the amount of information an attacker can learn from declassification statements.
For this, \tool{} directly checks that downgrades are bounded (\S \ref{sec:monad}) and it relies on the underlying security monad to ensure that the
adversary's knowledge remains constant, \ie there are no leaks, between two @downgrade@s.
As a result, the underlying security monad needs to enforce termination-sensitive non-interference.
Alternatively, one can use a monad enforcing termination-insensitive non-interference, such as LIO~\cite{stefan2011flexible}, and additionally prove termination, \eg using Liquid Haskell's termination checker.

%\mg{Add a forward pointer to the discussion section on security guarantees?}

\section{Refinement Types Encoding}
\label{sec:encoding}

We saw that our bounded downgrade function is correct, 
if each query is coupled with a function @approx@ that correctly 
computes the underapproximation of posterior knowledge. 
Here, we show how refinement types can specify correctness of @approx@, 
in a way that permits decidable refinement type checking. 
First (\S~\ref{subsec:encoding:abstractdom}),
we define the interface of abstract domains as a refined type class
that in~\S~\ref{subsec:encoding:approxes} we use to specify the abstractions of \indsets and knowledge. 
Next, we present two concrete instances of our abstract domains:
intervals (\S~\ref{subsec:encoding:intervals}) and powersets of intervals
(\S~\ref{subsec:encoding:powersets}).

\subsection{Abstract Domains}
\label{subsec:encoding:abstractdom}
\begin{figure}
\begin{mcode}
  class AbstractDomain $\adomtv$ $\cdomtv$ where 
    $\top$    :: $\adomtv$ <{\_ -> True , \_ -> False}>
    $\bot$    :: $\adomtv$ <{\_ -> False, \_ -> True }>
    $\in$    :: $\cdomtv$ -> $\adomtv$ -> Bool
    $\subseteq$    :: $\adomtv$ -> $\adomtv$ -> Bool
    $\cap$    :: d$_1$:$\adomtv$ <p$_1$, n$_1$> -> d$_2$:$\adomtv$ <p$_2$, n$_2$>
         -> {d$_3$:$\adomtv$ <p$_1\land$p$_2$, n$_1\lor$n$_2$> | d$_1 \subseteq$ d$_3$ && d$_2 \subseteq$ d$_3$}
    size :: $\adomtv$ -> {i:Int | 0 <= i}
    -- class laws 
    sizeLaw   :: d$_1$:$\adomtv$ -> {d$_2$:$\adomtv$ | d$_1 \subseteq$ d$_2$}
              -> {size d$_1 \leq$ size d$_2$}
    subsetLaw :: c:$\cdomtv$ -> d$_1$:$\adomtv$ -> {d$_2$:$\adomtv$ | d$_1 \subseteq$ d$_2$}
              -> {c $\in$ d$_1$ => c $\in$ d$_2$}
\end{mcode}
\caption{Abstract Domain Type Class}
\label{fig:encoding:types}
\end{figure}
Figure~\ref{fig:encoding:types} shows the @AbstractDomain $\adomtv$ $\cdomtv$@ 
refined type class interface stating that $\adomtv$ can abstract, \ie represent 
a set of values of, @$\cdomtv$@. 
For example, an instance @instance AbstractDomain $\adomi$ UserLoc@ 
states that the data type  $\adomi$  (that we will define in~\S~\ref{subsec:encoding:intervals})
abstracts @UserLoc@ (of~\S~\ref{subsec:downgrade}). 
The interface contains method definitions and  
class laws, and when required the abstract domain 
is indexed by abstract refinements. 

\paragraph{Class Methods}
The class contains six, standard, set---theoretic methods. 
Top ($\top$) and bottom ($\bot$), respectively represent 
the full and empty domains. 
Member $c \in d$ tests if the concrete value $c$ is included in the
  abstract domain $d$.
Subset $d_1 \subseteq d_2$ tests
  if the abstract domain $d_1$ is fully included in the abstract
  domain $d_2$.
Intersect $d_1 \cap d_2$  computes an
  abstract domain that includes all the concrete values that are
  included in both its input domains.
Finally, @size d@
computes the number of concrete values represented by an
  abstract domain.\looseness=-1

\paragraph{Class Laws}
We use refinement types to specify two class laws that should be satisfied 
by the $\subseteq$ and @size@ methods. 
@sizeLaw@
states that if @d$_1$@ is a subset of @d$_2$@, then the size of @d$_1$@
should be less than or equal to the size of @d$_2$@. @subsetLaw@ states that if @d$_1$@ is a
subset of @d$_2$@, then any concrete value in @d$_1$@ is also in @d$_2$@.
These methods have no computational meaning (\ie they return unit)
but should be instantiated by proof terms that satisfy the denoted laws. 
Even though we could have expressed more set-theoretic properties 
as laws, these two were the ones required to verify our applications.

\paragraph{Abstract Indexes}
In the types of top, bottom, and intersection, 
the type @$\adomtv$@ is indexed by two predicates @p@ and @n@ (both of type @$\cdomtv$ -> Bool@).
The positive predicate @p@ describes properties of concrete values that
are members of the abstract domain. Dually, the negative predicate @n@
describes properties of the values that do not belong to the abstract
domain. Intuitively, the meaning of these predicates is the following: 
\begin{mcode}
  $\adomtv$ <p,n> ~ {d:$\adomtv$ | $\forall$x. x$\in$d => p x &&  $\forall$x. x$\not\in$d => n x}
\end{mcode}
Yet, the right-hand side definition is using quantifiers which lead to undecidable 
verification. Instead, we used abstract refinements~\cite{abstract-refinements}
and the left-hand side encoding, to ensure decidable verification. 

The specification of the full domain 
@$\top$@
states that the positive predicate is @True@, \ie all elements belong to the domain, 
and the negative @False@, \ie no elements are outside of the domain.
Similarly, the empty domain @$\bot$@
has a @False@ positive predicate, \ie no elements are
in the domain, and @True@ negative predicate, \ie all elements
can be outside the domain.
Finally, the type signature for intersect @d$_1$ $\cap$ d$_2$@ returns a
domain @d$_3$@ whose positive predicate indicates it includes elements
included in @d$_1$@ and @d$_2$@ \ie @p$_1 \land$p$_2$@. The negative
predicate indicates points excluded from @d$_3$@ are points excluded
from either @d$_1$@ or @d$_2$@, \ie @n$_1 \lor$n$_2$@. The refinement on
@d$_3$@ ensure that @d$_3$@ is a subset $\subseteq$ of both @d$_1$@ and
@d$_2$@.
For abstract types in which these two predicates are omitted, 
the @\_ -> True@ predicate is assumed, which we will from now on abbreviate as 
@true@ and 
imposes no verification constraints. 

\subsection{Approximations of \indsets and knowledge}
\label{subsec:encoding:approxes}
In~Figure~\ref{fig:encoding:approx}, we use the positive and negative abstract indexes to 
encode the specifications of over- and under-approx\-i\-ma\-tions 
for \indsets and knowledge. 
We assume concrete types for 
$\adomtv$ and $\cdomtv$ with an 
@instance AbstractDomain $\adomtv$ $\cdomtv$@
and a query on the secret.
(In the previous sections for simplicity, we omitted the negative predicates and overapproximations.)

\begin{figure}
\begin{mcode}
  query :: $\cdomtv$ -> Bool 

  under_indset :: ($\adomtv$ <{\x ->   query x, true}>,
                   $\adomtv$ <{\x -> $\lnot$ query x, true}>)
  over_indset  :: ($\adomtv$ <{true, \x -> $\lnot$ query x}>,
                   $\adomtv$ <{true, \x ->   query x}>)

  underapprox :: p:$\adomtv$ -> 
    ($\adomtv$ <{\x ->   query x $\land$ (x $\in$ p), true}>,
     $\adomtv$ <{\x -> $\lnot$ query x $\land$ (x $\in$ p), true}>)
  underapprox p = (dT $\cap$ p,dF $\cap$ p) 
    where (dT,dF) = over_indset
  overapprox  :: p:$\adomtv$ ->
    ($\adomtv$ <{true, \x -> $\lnot$ query x $\lor$ (x $\not \in$ p)}>,
     $\adomtv$ <{true, \x ->   query x $\lor$ (x $\not \in$ p)}>)
  overapprox p = (dT $\cap$ p,dF $\cap$ p) 
    where (dT,dF) = over_indset
\end{mcode}
\caption{Specifications of Approximations 
for concrete $\adomtv$ and $\cdomtv$ that instantiate \texttt{AbstractDomain}.}
\label{fig:encoding:approx}
\end{figure}

\paragraph{Approximations of \indsets}
A query's \indsets  is a tuple whose first element is an abstract domain
that represents secrets that satisfying the query
and the second element is an abstract domain that represents secrets
that falsify the query.\looseness=-1

The specification of the \indsets @under_indset@
says the first domain only includes secrets for which the @query@ is
@True@ and the second domain only includes secrets for which the
@query@ is @False@ (the positive predicates). The negative predicates do
not impose any constraints on the elements that do not belong to the
domain. This means the domains can exclude any number of secrets, as
long as the secrets that are included are correct, \ie it is an
under-approximation.

Dually, the over-approximation @over_indset@ sets the negative predicate to
exclude all points for which the @query@ evaluates to @False@ for the
domain corresponding to the @True@ response and the second domain
(corresponding to the  @False@ response) excludes all points for which the @query@ evaluates to
@False@. The positive predicates are just true.
The domains can include any number of secrets as long as they are not
leaving out any secrets that are correct, \ie it is an
over-approximation.

\paragraph{Approximations of knowledge}
By combining the prior
knowledge of the attacker with the \indset for the query, we derive
an approximation of the attacker's knowledge after they observe the query.
Figure~\ref{fig:encoding:approx} shows the specifications for the
knowledge under-approximation @underapprox@ and the over-approximation
@overapprox@. @underapprox@ is similar to the type of @under_indset@,
except the positive predicate is strengthened to express that  
all the elements of the domain should also belong to the prior knowledge
@p@. Similarly, @overapprox@ specifies that the elements that do not
belong in the posterior knowledge, should neither be in the prior nor
the \indset. Each approximation is implemented by a pair-wise 
intersection with the respective \indsets and can be verified thanks 
to the precise type we gave to intersection. 

\paragraph{Precision}
The refinement types ensure our definitions are correct, 
but they do not reason about the precision of the abstract domains. 
For example, the bottom and top domains  
are vacuously correct solutions for under- and over-approximations, 
respectively. But, these domains are of little use as \indsets, since they ignore all
the query information. 
%In general a domain is more precise for an
%under-approximation if its size is higher, for an
%over-approximation if its size is lower.
It is unclear if precision can be encoded 
using refinement types. Instead, we empirically evaluate precision in  \S~\ref{sec:evaluation}.

\subsection{The Interval Abstract Domain}
\label{subsec:encoding:intervals}

Next we define \adomi, the interval abstract domain 
that can abstract any secret type $\cdom$, 
constructed as a product of integers (like the @UserLoc@ of~\S~\ref{sec:overview})
or types that can be encoded to integers (\eg booleans or enums).
\adomi is defined as follows:
\begin{mcode}
  -- $\cdom$ = Int $\times$ Int $\times$ ... 
  data $\adom$Int = $\adom$Int {lower :: Int, upper :: Int}
  type Proof p x = {v:$\cdom$<p> | v = x }

  data $\adomi$ <p::$\cdom$ -> Bool, n::$\cdom$ -> Bool> 
    = $\adomi$ { dom :: [$\adom$Int]
          , pos :: x:{$\cdom$| x $\in$ dom } -> Proof p x  
          , neg :: x:{$\cdom$| x $\not \in$ dom } -> Proof n x }
    | $\topi$ { pos :: x:$\cdom$ -> Proof p x }
    | $\boti$ { neg :: x:$\cdom$ -> Proof n x }
\end{mcode}

$\adomi$ has three constructors. $\topi$ and $\boti$ respectively denote
the complete and empty domains. @$\adomi$@ represents the domain of any
n-dimensional intervals, where @n@ is the length of @dom@. 
An interval @$\adom$Int@ 
represents integers between @lower@ and @upper@. For a secret
$\concval = \concval_1 \times \concval_2 \times \ldots \concval_n$, an
\adomi represents each @$\concval_i$@ 
by the @i@th element of its @dom@ (@$\concval_i \in$ (dom!i)@)
in the @n@ dimensional space. For example, 
@domEx = [($\adom$Int 188 212), ($\adom$Int 112 288)]@  
is the rectangle of $x \in [188, 212]$ and $y \in [112, 288]$ in the two
dimensional space of @UserLoc@.

\paragraph{Proof Terms}
The @pos@ and @neg@ components in the $\adomi$ definition 
are proof terms that give meaning to the positive @p@ and negative @n@
abstract refinements.
The complete domain \topi contains the proof field @pos@
that states that every secret \concval\ should satisfy the 
positive predicate @p@ (\ie @x: $\cdom$ -> Proof p x@) and the empty
domain contains only the proof @neg@ for the negative predicate @n@.
Due to syntactic restrictions that abstract refinements can only be
attached to a type for SMT-decidable
verification~\cite{abstract-refinements}, the proof terms are
encoded as functions that return the secret, while providing evidence
that the respective predicates are inhabited by possible secrets.
In \adomi this is encoded by setting preconditions to the
proof terms: the type of the @pos@ field states that each @s@ that
belongs to @dom@ should satisfy @p@, while the @neg@ field states that
each @x@ that does not belong to @dom@ should satisfy @n@. 
%(see \S~\ref{sec:overview} for an example \sng{put an example in
%\S~\ref{sec:overview}}).

When an $\adomi$ is constructed via its data constructors, 
the proof terms should be instantiated by explicit proof functions. 
For example, below we show that the @domEx@ (described above)
only represents elements that are @nearby (200,200)@.
\begin{mcode}
  example :: $\adomi$ <{\s -> nearby (200,200) s, true}>
  example = $\adomi$ domEx exPos (\x -> x)
  
  exPos :: s:{UserLoc | s $\in$ domEx } 
        -> {o:UserLoc | nearby (200,200) s && o = s} 
  exPos (UserLoc x y) = UserLoc x y 
\end{mcode}
The proof term @exPos@ is an identity function 
refined to satisfy the @pos@ specification. 
Once the type signature of @exPos@ is explicitly written, 
Liquid Haskell is able to automatically verify it. 
Automatic verification worked for all non-recursive queries, 
but for more sophisticated properties (\eg in the definition of the intersection function)
we used Liquid Haskell's theorem proving facilities~\cite{10.1145/3299711.3242756} 
to establish the proof terms. 
Importantly, when $\adomi$ is used opaquely 
(like in @approx@ in Figure~\ref{fig:encoding:approx}), 
the proof terms are automatically verified.\looseness=-1

\paragraph{AbstractDomain Instance}
We implemented the methods of the @AbstractDomain@ class for the 
\adomi data type as interval arithmetic functions lifted to
n-dimensions. $\in$ checks if any secret is between @lower@ and
@upper@ for every dimension. $\subseteq$ checks if the intervals
representing the first argument is included in the intervals
representing the second argument. $\cap$ computes a new list of
intervals to represent the abstract domain, that includes only the
common concrete values of the arguments. Size just computes the 
number of secrets in the domain, which can be interpreted as the domain's volume.
Our implementation consists of 360 lines of (Liquid) Haskell code, 
the vast majority of which constitutes explicit proof terms for 
@pos@ and @neg@ fields and the class law methods. 
By design, \adomi uses a list to abstract secrets 
that are sums of any number of elements, thus this class instance 
can be reused by an \tool user to abstract various secret types. 

\subsection{The Powersets of Intervals Abstract Domain}
\label{subsec:encoding:powersets}

To address the internal imprecision of the interval abstract domains, 
we follow the technique of~\cite{powerset1,powerset2} and define 
the powerset abstract domain \adomp, \ie a set of interval domains. 
Similar to intervals, the powerset
\adomp is also parameterized with the positive and negative predicates:
\begin{mcode}
data $\adomp$ <p::$\cdom$ -> Bool, n::$\cdom$ -> Bool> = $\adomp$ {
    dom$_i$ :: [$\adomi$] , dom$_o$ :: [$\adomi$]
  , pos :: x:{$\cdom$| x $\in$ dom$_i$ $\land$ x $\not\in$ dom$_o$} -> Proof p x  
  , neg :: x:{$\cdom$| x $\not\in$ dom$_i$ $\lor$ x $\in$ dom$_o$} -> Proof n x }
\end{mcode}
\adomp contains four fields. 
@dom$_i$@ is the set (represented as a list) of intervals
that are contained \emph{in} the powerset.
@dom$_o$@ is the set of intervals that are \emph{excluded} from the
powerset. This representation backed by two lists gives flexibility to
define powersets by writing regions that should be included and
excluded, without sacrificing generality or correctness (as guaranteed
by our proofs). Moreover, this encoding of the powerset makes our
synthesis algorithm simpler (\S~\ref{sec:synthesis}).
% To precisely represent domains -- especially over-$\!$ approximations,
% as we shall see next --  
% it is ofter convenient to \textit{exclude} intervals from the
% abstraction. 
The proof terms provide the boolean predicates that give semantics to
the secrets contained in the powerset, similar to the interval abstract domain
(\S~\ref{subsec:encoding:intervals}). We do not need a separate top
$\top$ and bottom $\bot$ for \adomp as they can be represented using
\topi or \boti in the @pos@ list.

\paragraph{AbstractDomain Instance}
We implemented the methods of the @AbstractDomain@ class for the 
powerset abstraction in 171 lines of code. 
A concrete value belongs to ($\in$) the powerset \adomp if it belongs
to any individual interval of the @dom$_i$@ list but not to any individual
interval of the @dom$_o$@ list. The subset @d$_1$ $\subseteq$ d$_2$@ operation checks
if each individual interval in the inclusion list @dom$_i$@ of @d$_1$@  is a
subset of at least one interval in the inclusion list @dom$_i$@ of @d$_2$@ and
also that none of the individual intervals in the exclusion list
@dom$_o$@ of @d$_1$@ is a subset of any interval in @dom$_o$@ of @d$_2$@. This operation
returns @True@ if the first powerset is a subset of the second, but
if it returns @False@ it may or may not be powerset. We have not found
this to be limiting in practice, as this criteria is sufficient for
verification. We plan to improve the accuracy via better algorithms 
in future work. Intersection @d$_1$ $\cap$ d$_2$@ produces a new powerset,
whose inclusion list is made of pairwise intersecting intervals from
@dom$_i$@ of @d$_1$@ and @dom$_i$@ of @d$_2$@ and the exclusion interval list is simply the
union of all intervals in the individual exclusion lists @dom$_o$@ of @d$_1$@ and
@dom$_o$@ of @d$_2$@. Size is the sum of the size of all intervals 
in the inclusion list minus the size of all intervals in the exclusion list.

\section{Synthesis of Optimal Domains}
\label{sec:synthesis}

We use synthesis in \tool to automatically generate \indsets
that satisfy the correctness types of Figure~\ref{fig:encoding:approx}
for each downgrade query. Our
synthesis technique proceeds in three steps: first, \tool extracts the
sketch of the posterior computation (\S~\ref{subsec:sketch}). Second, it
translates this to SMT constraints with relevant optimization directives
to synthesize the abstract domains (\S~\ref{subsec:smt-synth}). Finally,
the SMT synthesis is iterated to allow synthesis of
 powersets of any size (\S~\ref{subsec:synth-powerset}).
To efficiently perform these synthesis steps using SMT, 
we used a very restrictive form of the query language (\S~\ref{subsec:synthesis:queries}).

\subsection{The query language}
\label{subsec:synthesis:queries}
The queries analyzed by \tool 
are Haskell functions 
that take one input, of the secret type, 
and return a boolean: @query :: $\cdomtv$ -> Bool@ 
(as per Figure~\ref{fig:encoding:approx}).

For algorithm and efficient synthesis and verification, all the queries 
we tried are restricted to linear arithmetic, booleans, and data types
that have a direct, syntactic translation to 
SMT functions restricted to 
decidable logic fragments.  
Concretely, the queries can call other functions 
that belong to the same fragment,
but 
recursive definitions of queries are rejected by 
\tool. 

\paragraph{Supporting other query classes}
The query language can be easily extended to support non-boolean queries with finitely many outputs. 
This can be done 
% In
% practice, any return type that represents a finite set of outputs is
% supported 
by computing one \indset per possible output.
Further, our secrets currently and for simplicity are restricted to integer products,
but they can be easily extended to other domains with decidable
decision procedures (\eg datatypes).
Extensions to 
undecidable secret types (\eg floating points, strings) 
has unclear implications and is deferred 
to future work.

\subsection{Synthesis Sketch}
\label{subsec:sketch}
We use syntax-directed synthesis by starting with a 
sketch~\cite{solar2006combinatorial}, \ie a
partial program, for the \indsets based on their type specifications 
in Figure~\ref{fig:encoding:approx}.
For example, the sketch for the under-approximate \indsets would be:
\begin{mcode}
under_indset = ($\svar$::$\adom$ <{\x ->   query x, true}>,
                $\svar$::$\adom$ <{\x -> $\lnot$ query x, true}>)
\end{mcode}

Following the structure of the type we simply introduce
\textit{typed} holes of the form @$\svar$::$\tau$@ for each abstract domain,
which for this case is (refined) \adom.

\subsection{\textsc{Synth}: SMT-based Synthesis of Intervals}
\label{subsec:smt-synth}

We define the procedure \textsc{Synth} that given a
typed hole of an abstract domain, the number of fields in the secret $n$,  
and the kind of approximation (over or under), it 
returns a solution, \ie an abstract domain that satisfies the hole's type.
As an example, consider the below solution to first typed hole of @under_indset@. All @l@ and @u@ are symbolic integers.
\begin{mcode}
  $\svar$  :: $\adomi$ <{\x -> query x, \_ -> True}>
  $\svar$   = $\adomi$ dom pos neg
  dom = [$\adom$Int l$_{1}$ u$_{1}$, ..., $\adom$Int l$_{n}$ u$_{n}$]
\end{mcode}

The above solution is using the $\adomi$ applied to the domain list @dom@ and the @pos@ and @neg@ proof terms. 
The proof terms for our (non recursive) queries follow 
concrete patterns (as the example of~\S~\ref{subsec:encoding:intervals})
and are generated from syntactic templates.
The @dom@ is a list of ranges @$\adom$Int@ that contains 
symbolic integers as lower (@l$_i$@) and upper (@u$_i$@) bounds, while the length $n$ of the list  
is the number of fields of the secret data type. 

To find concrete values for the symbolic integers @l$_i$@ and @u$_i$@, \textsc{Synth} mechanically generates 
SMT implications based on the type indexes.
Since the positive index states that all elements @x@ on the domain should 
satisfy @query x@ and the negative index states that all elements outside of the domain 
should satisfy @True@, the following SMT constraint is mechanically generated: 
\begin{mcode}
  $\forall$ x. (x$\in$ dom =>  query x) && (x$\not\in$ dom => True)
\end{mcode}

Such constraints (see~\S~\ref{subsec:overview:anosy} for a concrete example)
are sent to the SMT, by a direct, syntactic translation 
of the Haskell instance method $\in$
and the @query@ definitions into Z3 functions (\S~\ref{subsec:synthesis:queries}).  
%Additionally, the correctness conditions for the \indset depends on the
%query definition. \tool then lifts the @query@ definition to the SMT-decidable 
%fragment of logic that includes integer arithmetic and
%booleans. Queries take a secrets, \ie a tuple of integers, and return a
%boolean, making this fragment of logic sufficient to express operations
%in it.

Solving such constraints gives us a value for @dom@ 
if a solution exists. In practice,
however, such solutions are often just a point, \ie the abstract domain
contains only one secret. Although this is a correct solution, it is not
precise. 
%We would prefer the largest posterior if
%we are under-approximating, and the smallest if we are
%over-approximating. 
To increase precision we add optimization directives to constraints
depending on the type of our approximation. 
That is, for $i\in\{1\dots n\}$  we add 
@maximize u$_{i}$ - l$_{i}$@ or @minimize u$_{i}$ - l$_{i}$@ for
under-approximations and over-approximations respectively.
%
%\begin{mcode}
%  maximize u$_{bi}$ - l$_{bi}$ $\quad\textnormal{for under- or}\quad$  minimize u$_{bi}$ - l$_{bi}$ $\textnormal{for over-approx}\quad$ 
%\end{mcode}
These optimization constraints are handed to  an SMT solver
that supports optimization directives~\cite{bjorner2015nuz} and the
produced model is an intended solution for @dom@. We
used the Pareto optimizer of Z3~\cite{bjorner2015nuz}, such that no
single optimization objective dominates the solution. For example, if
two domains of sizes $400 \times 1$ and $20 \times 20$ are valid solutions, \tool will prefer
the latter.

\subsection{\textsc{IterSynth}: Iterative Synthesis of PowerSets}
\label{subsec:synth-powerset}

\begin{algorithm}[t]
\small
\caption{Iterative Synthesis of Powersets}
\label{alg:iter-synth}
\begin{algorithmic}[1]
\Procedure{IterSynth}{k, $n$, $\tau$, apx}
\State{\texttt{dom\_i} $\gets$ [\textsc{Synth} (\adomp \texttt{[$\square$]} \texttt{[]} \_ \_)::$\tau$ n apx]}
\State{\texttt{dom\_o} $\gets$ \texttt{[]}}
\For{i = 2 to k}
\If{apx == under}
\State{\texttt{dom\_t} $\gets$ \textsc{Synth} (\adomp (\texttt{dom\_i} ++ \svar) \texttt{dom\_o} \_ \_)::$\tau$ n apx}
\State{\texttt{dom\_i} $\gets$ \texttt{dom\_i} ++ [\texttt{dom\_t}]}
\Else
\State{\texttt{dom\_t} $\gets$ \textsc{Synth} (\adomp \texttt{dom\_i} (\texttt{dom\_o} ++ \svar) \_ \_)::$\tau$ n apx}
\State{\texttt{dom\_o} $\gets$ \texttt{dom\_o} ++ [\texttt{dom\_t}]}
\EndIf
\EndFor
\State \Return (\adomp \texttt{dom\_i} \texttt{dom\_o} \_ \_)
\EndProcedure
\end{algorithmic}
\end{algorithm}

Powerset abstract domains (\S~\ref{subsec:encoding:powersets})
are synthesized by Algorithm~\ref{alg:iter-synth} 
that iteratively increments the powersets with 
individual intervals to avoid 
scalability problems faced by Z3 when optimizing multiple intervals
at once.

The algorithm takes as arguments the  number of intervals @k@ to be
included in the powerset, the number of fields in the secret $n$, the
refinement type of the powerset domain $\tau$, and the kind of approximation @apx@
 (@under@ or @over@). It first runs 
\textsc{Synth} (\S~\ref{subsec:smt-synth}) to 
generate the first interval, with the top level type properly propagated to the hole.
If this is for an under-approximation, more
such intervals can be added to the powerset to boost the precision.
Conversely, if the first synthesized interval is an over-approximation,
then more intervals can be eliminated from the powerset to return a more
precise over-approximation. At each iteration, the algorithm creates a
new placeholder interval \svar and \textsc{Synth} solves it,
incrementally building up the inclusion list @dom_i@, or the exclusion
list @dom_o@. Finally, the powerset is returned after $k$ iterations.
This is \tool's general synthesis algorithm since for $k = 1$ the returned
powerset has a single interval. 

As a final step, the
returned powerset is lifted to the Haskell source and substituted in the
sketch in \S~\ref{subsec:sketch}, which as a sanity check is validated by Liquid Haskell.

\paragraph{Discussion}

Traditional abstract interpretation based techniques will
refine the domains, as the query is evaluated with small step semantics,
leading to imprecision at each step. In contrast, \tool is more 
precise (as we show in \S~\ref{sec:evaluation}), because the final
abstract domain is synthesized in the final step after accumulating
constraints. However, Z3 does not give precise solutions when there are
too many maximize/minimize directives (more than 6 in our experience)
and it does not handle non-linear objectives well. We leave exploration of
better optimization algorithms to future work.

\section{Evaluation}
\label{sec:evaluation}

We empirically evaluated \tool's performance using two case
studies. 
In the first one (\S~\ref{subsec:evaluation:verif-perf}), we analyze efficiency and precision of \tool{} when verifying and synthesizing \indsets using a set of micro-benchmarks from prior work.
In the second one (\S~\ref{subsec:evaluation:monad}), we use the \tool{} monad to construct an application that performs multiple queries (similar to those of \S~\ref{sec:overview}) while enforcing a security policy on the attacker's knowledge.
With this case study, we evaluate how losses of precision introduced by \tool{}'s abstract domains affect the ability of answering multiple queries.

% the attacker's knowledge evolves across multiple queries and (2) to
% enforce a simple security policy over such knowledge. The goal of this
% case study is evaluating how losses of precision introduced by \tool{}'s
% approximation affect the ability of answering multiple queries.

% efficiency and precision

% We pose the following questions in our evaluation:

% \begin{itemize}
%   \item Can \tool verify and synthesize posteriors with different
%   abstract domains on a set of problems previously discussed in
%   literature? 
%   \item Can the \tool monad be used to write an application that can
%   process queries while guaranteeing bounds on information leakage?
%   (\S~\ref{subsec:evaluation:monad})
% \end{itemize}

% \mg{I'd rather stress what we are measuring in each case study}

\paragraph{Experimental setup}

\tool is a GHC plugin built against GHC 8.10.1. All refinement types
were verified with LiquidHaskell 0.8.10. Z3 4.8.10 was used to
synthesize the bounds of the abstract domains. All experiments were
performed on a Macbook Pro 2017 with 2.3 GHz Intel Core i5 and 8GB RAM.\looseness=-1

% We start by describing the experimental setup, and then present
% the programs we use in our case studies
% (\S~\ref{subsec:benchmark-progs}). We continue with our first case
% study, where we analyze \tool's 
% performance using micro-benchmarks. The goal of this case study is to
% evaluate \tool{}'s precision and efficiency (in terms of both
% verification and synthesis time) on selected examples. Finally, we
% describe our second case study, where we use \tool (1) to monitor how
% the attacker's knowledge evolves across multiple queries and (2) to
% enforce a simple security policy over such knowledge. The goal of this
% case study is evaluating how losses of precision introduced by \tool{}'s
% approximation affect the ability of answering multiple queries.

\subsection{Verification \& Synthesis of \indsets} % of under-/over-approximate domains}
% Case study: Microbenchmarks on \indsets Synthesis \& Verification
\label{subsec:evaluation:verif-perf}

In this case study, we analyze the \tool's performance with respect to the verification and synthesis of \indsets.

\begin{figure*}
  \begin{subfigure}{\textwidth}
    \centering
    \small
    \begin{tabular}{|r|r|r|r|r|r|r|r|r|r|r|}
      \hline
        & \multicolumn{4}{c|}{\thead{Under-approximation}} & \multicolumn{4}{c|}{\thead{Over-approximation}} \\  
        \cline{2-9}
        \thead{\#} & \thead{Size} & \thead{\% diff.} & \thead{Verif. time} & \thead{Synth. time} & \thead{Size} & \thead{\% diff.} & \thead{Verif. time} & \thead{Synth. time} \\  
      \hline
      B1 & 259 / 9620          & 0 / 27    & 2.78 $\pm$ 0.03 & 1.11  $\pm$ 0.01 & 259 / 13505         & 0 / 2     & 2.64 $\pm$ 0.03 & 1.07  $\pm$ 0.01 \\
      B2 & 2.21e+05 / 1.01e+07 & 78 / 58   & 3.62 $\pm$ 0.02 & 9.26  $\pm$ 0.04 & 2.02e+06 / 2.54e+07 & 100 / 5   & 3.17 $\pm$ 0.02 & 4.00  $\pm$ 0.12 \\
      B3 & 4 / 664             & 0 / 25    & 3.12 $\pm$ 0.06 & 0.90  $\pm$ 0.07 & 4 / 888             & 0 / 0     & 2.83 $\pm$ 0.03 & 0.90  $\pm$ 0.01 \\
      B4 & 3.53e+04 / 1.35e+05 & 100 / 100 & 3.66 $\pm$ 0.04 & 20.92 $\pm$ 0.11 & 9.22e+12 / 2.81e+13 & 67200 / 0 & 3.29 $\pm$ 0.08 & 10.87 $\pm$ 0.01 \\
      B5 & 360 / 5.04e+06      & 83 / 25   & 3.81 $\pm$ 0.04 & 1.38  $\pm$ 0.04 & 35460 / 6.72e+06    & 1542 / 0  & 3.47 $\pm$ 0.04 & 0.89  $\pm$ 0.01 \\
      \hline
    \end{tabular}
    \caption{\small Interval abstract domain}\label{table:results:interval}
  \end{subfigure}

  \ 

  \begin{subfigure}{\textwidth}
    \centering
    \small
    \begin{tabular}{|r|r|r|r|r|r|r|r|r|r|r|}
      \hline
        & \multicolumn{4}{c|}{\thead{Under-approximation}} & \multicolumn{4}{c|}{\thead{Over-approximation}} \\  
        \cline{2-9}
        \thead{\#} & \thead{Size} & \thead{\% diff.} & \thead{Verif. time} & \thead{Synth. time} & \thead{Size} & \thead{\% diff.} & \thead{Verif. time} & \thead{Synth. time} \\  
      \hline
      B1 & 259 / 13246         & 0 / 0     & 4.51 $\pm$ 0.05 & 1.13  $\pm$ 0.02 & 259 / 13505         & 0 / 2     & 4.34 $\pm$ 0.03 & 1.08  $\pm$ 0.01 \\
      B2 & 6.78e+05 / 1.62e+07 & 33 / 33   & 5.32 $\pm$ 0.09 & 14.34 $\pm$ 0.11 & 1.80e+06 / 2.54e+07 & 78 / 5    & 5.17 $\pm$ 0.02 & 4.89  $\pm$ 0.09 \\
      B3 & 4 / 880             & 0 / 0     & 5.29 $\pm$ 0.09 & 1.07  $\pm$ 0.03 & 4 / 888             & 0 / 0     & 4.99 $\pm$ 0.03 & 1.03  $\pm$ 0.01 \\
      B4 & 3.88e+05 / 4.00e+05 & 100 / 100 & 5.78 $\pm$ 0.03 & 54.89 $\pm$ 0.23 & 9.22e+12 / 2.81e+13 & 67200 / 0 & 5.48 $\pm$ 0.08 & 30.57 $\pm$ 0.07 \\
      B5 & 720 / 6.70e+06      & 67 / 0    & 6.02 $\pm$ 0.07 & 13.26 $\pm$ 0.09 & 6300 / 6.72e+06     & 192 / 0   & 5.96 $\pm$ 0.04 & 15.25 $\pm$ 0.03 \\
      \hline
    \end{tabular}
    \caption{\small Powerset of intervals with size 3} \label{table:powerset-results}
  \end{subfigure}
% \todo{I'd suggest adding columns that express precision as percentage of covered volume over the original cardinality from Table 1. }
  \caption{\small Ind. sets synthesis and verification of posteriors. Column
  \emph{Size} reports the size of the synthesized
  \indsets, where $x$ is the size of the \texttt{True} set and $y$
  of the \texttt{False} set in $x/y$. \emph{\% diff} shows the percentage
  difference of the size from precise \indset in
  Table~\ref{table:results:concrete} (lower value is 
  better). \emph{Verif. time} and \emph{Synth. time} columns report
  (in seconds) the median and the semi-interquartile  over 11 runs.}
  \label{fig:results:case-study1}
\end{figure*}

\begin{table}
  \centering
  \small
  \caption{\small Number of fields in the secret, and size of the
  precise \indsets $x / y$ for our benchmarks, where $x$ and $y$ denotes
  the number of secrets that evaluate to \texttt{True} and
  \texttt{False}, respectively.}
  \label{table:results:concrete}
  \begin{tabular}{|r|r|r|r|}
    \hline
    \thead{\#} & \thead{Name} & \thead{No. of fields} & \thead{Size of \indsets} \\  
    \hline
    B1 & Birthday & 2 & 259 / 13246         \\
    B2 & Ship     & 3 & 1.01e+06 / 2.43e+07 \\
    B3 & Photo    & 3 & 4 / 884             \\
    B4 & Pizza    & 4 & 1.37e+10 / 2.81e+13 \\
    B5 & Travel   & 4 & 2160 / 6.72e+06     \\
    \hline
  \end{tabular}
\end{table}

\paragraph{Benchmark Programs}
Our benchmarks consist of 5 problems from
\citet{mardziel-beliefpol-2013}, which represent a diverse set of
queries (\textsc{B3} and \textsc{B4} come from a targeted advertisement
case study from Facebook~\cite{facebook-travel}). We selected these
benchmarks to illustrate that \tool supports similar classes of queries
as existing prior work and to compare performance and precision with
available tools.
% (with a few minor changes): 

\begin{asparaitem}
\item[(\textsc{B1})] \emph{Birthday} checks if a user's birthday, the secret, is within
the next 7 days of a fixed day\footnote{We only use
the deterministic version of the \emph{Birthday} problem.}.

\item[(\textsc{B2})] \emph{Ship} calculates if a ship can aid an island
based on the island's location and the ship's onboard capacity.\looseness=-1

\item[(\textsc{B3})] \emph{Photo} checks if a user would be possibly interested in a wedding
photography service by checking if they are female, engaged, and in a
certain age range.
% This is an encoding of a real targeted advertisement used by Facebook.

\item[(\textsc{B4})] \emph{Pizza} checks if a user might be interested in ads of a local pizza
parlor, based on their birth year, the level of school attended, their address latitude and longitude (scaled by $10^6$).\looseness=-1
% The query checks if
% the user is in a certain region and whether they match age and education
% criteria associated with pizza eating habits.

\item[(\textsc{B5})] \emph{Travel} tests for a user interest in travels by
checking if the user speaks English, has completed a high level of
education, lives in one of several countries, and is older than 21.\looseness=-1
% This query is adapted from a Facebook case study based on a campaign run
% by a tourism agency.

% This query calculates the Manhattan distance (an example of a
% relational
% query) of a ship from some concrete location
% and if the ship can carry a certain number of people onboard.
\end{asparaitem}

% We believe these problems~\cite{mardziel-beliefpol-2013} represent a
% diverse set of queries from real applications that can be meaningfully
% protected with quantitative information leakage policies. \emph{Photo} and
% \emph{Travel} are examples taken from a targeted advertisement case
% study from Facebook~\cite{facebook-travel}.
% \mg{I don't like this paragraph too much. It sounds a bit defensive. Moreover, I'm not sure whetehr these are really diverse. }

For each problem, we encode the query as a Haskell function with the appropriate refinement type~\cite{vazou2017refinement} where the secret domain is represented as a Haskell datatype for which we use the same bounds as \citep{mardziel-beliefpol-2013}.
%
% For the secret domain, we use the same bounds as \citep{mardziel-beliefpol-2013}.
%
% \mg{What does ``the function reflected into the refinement type~\cite{vazou2017refinement}'' mean?}
%
Table~\ref{table:results:concrete} reports the number of fields in the secret, and
the size of the precise \indsets for each benchmark as $x / y$, where
$x$ denotes the size of the precise \indset for the @True@ response from
query and $y$ is the size when the query responds @False@. 
%

% We will
% demonstrate that precise \indsets are not always 
% representable in a specific abstract domain without loss of precision
% (\S~\ref{subsec:evaluation:verif-perf}), necessitating the need for
% under-/over-approximations.

% \mg{Where are we \emph{demonstrating} that precise \indsets are not representable?}

% \mg{Better names for table columns?}

% The secrets for these problems were encoded as Haskell datatypes.
% After translating these queries to Haskell, all the annotations to
% specify bounds and to generate under-/over-approximate domains were
% provided. The bounds were taken as-is without any modification from
% \citet{mardziel-beliefpol-2013}. The abstract domains were synthesized
% during the compilation process by adding the necessary sound/complete
% annotations and then verified by LiquidHaskell.

% The \emph{Dims} column provides the number of dimensions in the secret.
% Each dimension is represented as a field in the secret datatype. The
% \emph{Actual} column lists the size of the indistinguishable set when
% the query returns {\sf True} and {\sf False} respectively. Note
% that this is the most precise representation of the query's
% indistinguishable set and it may or may not be represented exactly in an
% abstract domain.

\paragraph{Experiment}
For each benchmark, we use \tool{} to (1) synthesize the under- and
over-approximated \indsets for both results @True@ and @False@ and (2) verify that
the synthesized approximations match the refinement types
from~\S~\ref{sec:encoding}. We run each benchmark 11 times to collect synthesis
and verification times. We use a 10 second timeout for each Z3 call. The goal is
to evaluate the precision of the synthesized \indset and time taken for
synthesis and verification to run.

\paragraph{Intervals} % Results

Figure~\ref{table:results:interval} reports the results of our
experiments for both the under- and over-approximated \indsets using the
interval abstract domain.
Specifically, the column \emph{Size} reports the number of secrets in the
approximated \indset, the column \textit{Verif. time} reports the time (in
seconds) LiquidHaskell takes to verify the posteriors, and
the column \textit{Synth. time} reports the time (in seconds) taken for
synthesizing the approximate \indsets. The \emph{\% diff.} column 
lists the difference in size of the approximate \indsets with the
exact ones from Table~\ref{table:results:concrete}. The lower the
\emph{\% diff.} column value, the more precise is  the synthesized
\indset, \ie it is closer to the ground truth.\looseness=-1

For all our benchmarks, LiquidHaskell  quickly verifies the
correctness of the posteriors, in less than 4 seconds on
average. In some cases, like \textsf{B1} and \textsf{B3}, \tool can
synthesize the exact \indset for the @True@ result using a single
interval (for both approximations). For the @False@ set, however, the
tool returns an approximated result 
because the precise \indset is not representable using intervals.
%
% \mg{Any intuition for why this is the case?}

In 7 out of 10 synthesis problems, \tool synthesizes the
approximations in less than 5 seconds. The three outliers are the
synthesis of under-approximations for \textsf{B2} and the
synthesis of both approximations for \textsf{B4}.
\textsc{B2} uses a
relational query that creates a dependency between two secret fields, where the multi-objective maximization employed by Z3 runs
longer.
\textsc{B4} uses very large bounds (in the orders of $10^8$) which result in Z3 quickly finding a sub-optimal model but timing-out before finding an optimal solution.

\paragraph{Powersets of intervals}

Figure~\ref{table:powerset-results} reports the results of our
experiments using the powersets domain with 3 intervals. A higher number
gives more precision for representation of the \indset at the cost of
taking more time for synthesis, due to our iterative synthesis algorithm (\S~\ref{subsec:synth-powerset}).
%The data is presented in a similar format as the previous table.

For under-approximations, \tool{} successfully synthesizes both exact
\indsets for \textsf{B1} using powersets, even though
the @False@ set was not representable using just a single interval. 
For \textsf{B2} and \textsf{B3}, the powersets
significantly improve precision, \ie we synthesize larger
under-approximations.

This can be seen by comparing the \emph{\% diff.}
column between Figure~\ref{table:results:interval} and
\ref{table:powerset-results}, where the latter reports lower percentage
differences from ground truth.
In fact, for \textsf{B3}, \tool{} can almost
synthesize the entire \indset for @False@ with powersets of size 3 and
it can synthesize the exact \indset with powersets of size  4 (not
shown in Figure~\ref{table:powerset-results}). For \textsf{B4}, powersets only marginally improve precision due to SMT
optimization timing out. 
For over-approximations, we observe a similar increase in precision, in
particular in \textsf{B3} and \textsf{B5} where the
synthesized approximations are close to the exact values. 
\textsf{B4} slows down drastically because
synthesis of each interval takes almost 10 seconds due to SMT timeouts.

\paragraph{Discussion}

\tool synthesizes \indsets, and a function to compute a
posterior for any prior, incurring one-time cost for synthesis but
making posterior computation free at runtime. In contrast, prior tools
like @Prob@~\cite{mardziel-beliefpol-2013} need to run an expensive
static analysis each time when computing the posterior knowledge. While
the synthesis takes 54.2x longer on average than running Prob each time,
this cost is amortized over multiples runs of the program with \tool.

Moreover, \tool is more precise than @Prob@, as demonstrated by
difference from ground truth in benchmarks like \textsf{B3}
(Figure~\ref{table:powerset-results}). A difference of 0 indicates that \tool
synthesized an exact \indset. In contrast @Prob@'s
belief was 0.1429 (\ie had some uncertainty; 0 is exact) for the same
example in same conditions. \tool is more precise because it can
automatically split regions into intervals
(\S~\ref{subsec:synth-powerset}) whose union in the powerset
gives a better accuracy. 
For instance, in Figure~\ref{table:powerset-results}, a powerset of size 3 is enough to synthesize the exact \indset (\emph{\% diff.} is 0) for several benchmarks.

In our experience, iterative synthesis (\S~\ref{subsec:synth-powerset})
works better than existing techniques~\cite{backesleak09,
mardziel-beliefpol-2013} for queries 
(benchmarks \textsf{B1}, \textsf{B3}, and \textsf{B5}) that
contain point-wise comparisons, \ie the query checks if a secret $x$ is
one of several constant values $c_1, c_2, \ldots$, or in other words, formulas of the
form $x = c_1 \lor x = c_2 \lor \dots$.
These queries split the indistinguishable sets into a union of disjoint sets,
and the SMT solver efficiently identified the best possible solution
for the abstract domain. However, benchmarks that do not use point-wise
comparisons (like \textsf{B2}) perform equivalent to prior
work~\cite{mardziel-beliefpol-2013} in our experience.

\subsection{Secure Advertising System}
\label{subsec:evaluation:monad}

In this case study, we go back to the advertisement example in
\S~\ref{sec:overview} which we implement  using \tool{} to restrict the
information leaked through @downgrade@. The goal of this case study is
evaluating how the choice of abstract domains affects the number of declassification queries authorized by \tool{}.\looseness=-1

\begin{figure}
  \includegraphics[scale=0.4]{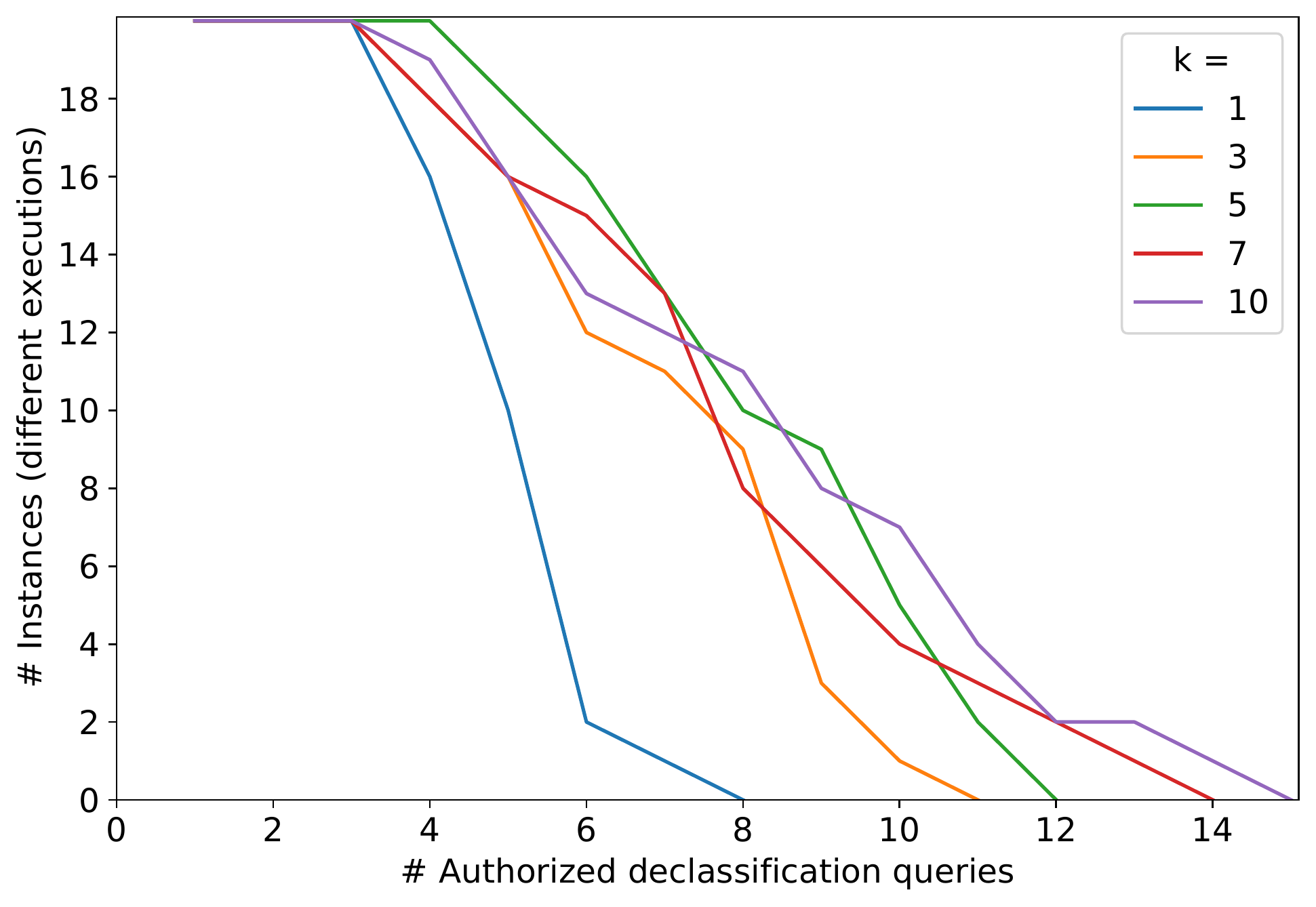}
  \caption{\small The lines show the number of execution instances
  (Y-axis) that were authorized for the $i$-th declassification query (X-axis).
  Each line corresponds to the under-approximated \indset of powersets of size $k$.}
  \label{fig:autoquery}
\end{figure}

\paragraph{Application}
We implemented the advertising query system from \S~\ref{sec:overview} in Haskell
using the @AnosyT@ monad, with the @UserLoc@ type as the secret.
The system executes a sequence of 50 queries (one per restaurant branch): % as queries 
we use the @nearby@ query from \S~\ref{sec:overview} with
the origin, denoting in this experiment the location of the restaurant, being a randomly generated point in the
$400 \times 400$ space. % rather than the constant (200, 200).

% corresponding to each location of the restaurant branch. Each query looks like
% the @nearby@ query from Figure~\ref{fig:userloc-defn}, except the location of
% the restaurant a randomly generated coordinate in the $400 \times 400$ space,
% instead of a constant (200, 200) coordinate. We have upto 50 such restaurants
% randomly generated in the $400 \times 400$ space.

\paragraph{Security policy and enforcement}
Our program implements the security policy @qpolicy@ from
\S~\ref{sec:overview}, which restricts the restaurant chain from
learning the user location below a set of 100 possible locations.
To easily enforce the security policy, we wrapped the advertising query in the
@downgrade@ operation of @AnosyT@ as in~\S~\ref{sec:monad}.
% just like the @nearby@ query in the
% @showAdNear@ function.

Initially, our system starts with a prior knowledge equivalent to the entire
secret domain $400 \times 400$ (i.e., the attacker does not have any information
about the secret). As the system executes queries, the @AnosyT@ monad tracks an
under-approximation of the attacker's posterior knowledge based on the query
result and on the prior. If the posterior complies with the policy, then the
monad outputs the query result and the system continues with the next
query. If a policy violation is detected, 
% the monad returns an error and
the system terminates the execution.

\paragraph{Experiment}
For each experiment, we generate a new user location randomly, used as
the secret, in the $400 \times 400$ space, and we run through the 50 queries for
every restaurant location. For each execution, we measure after how many
queries the system stops due to a policy violation. We repeat this experiment 20
times, to get the mean and standard deviation of query count and discuss them
below.

% We compute (1) the size of the knowledge tracked by the @AnosyT@ monad 
% location changes with the evaluation of each query, and when the system stops
% evaluating queries due to a policy violation. We repeat this experiment 20
% times, to get the mean and standard deviation of these metrics and discuss them
% below.

% We generate 20 input tuples (each consisting of a secret, a sequence of
% queries, and a policy) for the security monitor, we run the monitor for
% each input tuple, and we count the number of queries that the monitor
% authorizes before detecting a policy violation. 
% We perform the above process using \tool to generate both under- and
% over-approximations of the knowledge and we treat the number of
% authorized queries as a proxy for how fast the loss of precision
% degrades the ability to answer queries. 

\paragraph{Results}

Figure~\ref{fig:autoquery} reports the results of our experiments.
The line for each $k$, \ie the number of synthesized intervals in
the powerset, depicts the number of experiment instances that are
still running (Y-axis) after executing the $i$-th query (X-axis).
For example, in the $k=1$ powerset (equivalent to an interval), the
system was able to answer the first 3 queries in all 20 instances
without violating the policy, but only 2 instances were able to answer
the 6th query.

% We
% report the median and semi-interquartile ranges of the 20 runs. As we
% run more queries, the synthesized posterior could become imprecise (\ie
% not have the secret), thus that sequence of queries is aborted. We
% report in the dashed line how many of the 20 runs ran till the $i^{th}$
% step (X-axis).

% \todo{Update description once we have the new figure}
%
% Using the interval abstract domain, in all 20 runs the system could
% answer a maximum of 6 queries before detecting a policy violation (due to
% precision losses), with only 2 runs reaching the 6th query
% (Figure~\ref{subfig:k1under-autoquery}). 
% This is also reflected in the mean size of the
% under-approximation decreasing very rapidly. 

As the size of powersets increases (from 3 to 10), the system can compute
more precise under-approximations and, therefore, securely answer more
queries, as can be seen in the figure. Specifically, for powerset of size $k = 3$, 
the system answers
a maximum of 10 queries over 20 runs, with only 1 run reaching the 10th
query.
Similarly, the maximum number of queries answered increases to 14, due to
increased precision by using powersets of size 10.
Moreover, more than 10 instances answer more than 6 queries if
the size of powersets goes above 3. This shows that \tool can be used to build
a system, that can answer multiple queries sequentially with precision
without violating the declassification policy.

Figure~\ref{fig:autoquery} shows a tradeoff between number of queries
answered and the precision of the powersets. Higher sized powersets
($k=7$ or $10$) under-perform in the intermediate declassifcations from
5 to 7 (on the X-axis) when compared to $k=5$. The intersection of
powersets made of $k_1$ and $k_2$ intervals produces a powerset of
$k_1k_2$ intervals, of which many intervals are small or empty (as
individual powersets might have very little overlap). Hence a slightly
more imprecise powerset $k=5$ declassifies allows more instances of the
query to run. However, over a longer sequence of queries a higher sized
powerset performs better due to improved precision in tracking knowledge
(as can be seen from $k=10$ allowing 14 declassifcations).

\section{Related Work}
\label{sec:related}

\paragraph{Information-flow control}
% \paragraph{Information-flow control}
%\textit{Information-flow control:}
%
Language-based information-flow control (IFC)~\cite{SabelfeldM03} provides
principled foundations for reasoning about program security.
Researchers have proposed many enforcement mechanisms for IFC like {type systems}~\cite{broberg2017paragon, arden2012sharing, pottier2002information,li2006encoding,devriese2011information,russo2015functional, Polikarpovalifty}, static analyses~\cite{10.1145/2737924.2737957}, and runtime monitors~\cite{guarnieri2019information} to verify and enforce security properties like non-interference.
The \indsets and knowledge approximations computed by \tool can be used as a building block to enforce both  non-interference as well as more complex security policies, as we discuss below. 

% \paragraph{Computing indistinguishability relations}
% %
% Many IFC approaches~\cite{Clark05qif,backesleak09,KopfR10} explicitly compute a representation of the indistinguishability equivalence relation induced by programs, where initial program states are related only if an attacker cannot distinguish them through observations. 
% %
% Rather than using equivalence relations, we use a representation based on the partition induced by the indistinguishability relation,  where each \indset is one of the relation's equivalence classes. 

% \citet{Clark05qif} provide techniques to approximate the indistinguishability relation for straight line programs.
% %
% \citet{backesleak09} automates the synthesis of such equivalence relations using program verification techniques, and \citet{KopfR10} further improves the approach by combining it with sampling-based techniques.
% %
% Similarly to~\cite{backesleak09}, we automatically synthesize \indsets from programs. 
% %
% In contrast to~\cite{Clark05qif,backesleak09,KopfR10}, however, our the correctness of our \indsets is automatically machine-checked using refinement types.

%\textit{Use of knowledge in IFC:}
\paragraph{Use of knowledge in IFC}
% \paragraph{Use of knowledge in IFC}
%
The notion of attacker knowledge has been originally introduced to reason about dynamic IFC policies, where the notion of ``public'' and ``secret'' information can vary during the computation~\cite{askarov2007gradual,guarnieri2019information, van2015very}. %the effects of declassifying information~\cite{askarov2007gradual}.
%
% More generally, knowledge-based notions have been used to reason about . 
% A natural extension of the notion of knowledge is that of \emph{belief}, which is a knowledge, \ie set of possible secret values, equipped with a probability distribution describing how likely each secret is.
The notion of \emph{belief} consists of a knowledge, \ie set of possible secret values, equipped with a probability distribution describing how likely each secret is.
Existing
approaches~\cite{SweetTS0M18,mardziel-beliefpol-2013,guarnieri2017securing,kucera2017synthesis}
can enforce security policies involving probabilistic statements over an
attacker's belief, \eg ``an attacker cannot learn that a secret holds
with probability higher than 0.7''. We plan to deal with probability
distributions in future work. However, \tool synthesizes a function that
computes the posterior given a prior, eliminating the need to run the
full static analysis for each query execution. This enables applications
to directly use knowledge based policies without expensive static
analysis at runtime. Additionally \tool's posterior knowledge is
correct-by-construction and mechanically verified using LiquidHaskell's
refinement types, unlike existing tools~\cite{mardziel-beliefpol-2013}
which rely on (often complex) pen-and-paper proofs.
%
% These approaches have limited scalability.
%
% For instance, for the \emph{Birthday} example (\S\ref{sec:evaluation}), \tool{} computes the knowledge in $\sim$1 sec, whereas the analysis from~\cite{mardziel-beliefpol-2013} computes the probabilistic belief in  $\sim$17 sec.

Quantitative Information Flow approaches provide quantitative metrics, \eg{} {Shannon
entropy}~\citep{shannon2001mathematical}, {Bayes
vulnerability}~\citep{smith2009foundations}, and {guessing
entropy}~\citep{massey1994guessing}, that summarize the amount of leaked information.
For this, several approaches~\cite{Clark05qif,backesleak09,KopfR10} first compute a representation of a program's indistinguishability equivalence relation, whereas we represent the partition induced by the indistinguishability relation,  where each \indset is one of the relation's equivalence classes. 
% where related program states cannot be distinguished by attackers,  which is later used to derive quantitative bounds on the leaked information.
% %
% Rather than using equivalence relations, we represent the partition induced by the indistinguishability relation,  where each \indset is one of the relation's equivalence classes. 

% We remark that, if the prior is the entire domain, then the knowledge computed by \tool{} is (an approximation of) the \indset, so \tool{} can also be used to derive approximations of a program indistinguishability relation.
%
There are several approaches for approximating the indistinguishability relation in the literature. 
 \citet{Clark05qif} provide techniques to approximate the indistinguishability relation for straight line programs.
 \citet{backesleak09} automates the synthesis of such equivalence relations using program verification techniques, and \citet{KopfR10} further improve the approach by combining it with sampling-based techniques.
 Similarly to~\cite{backesleak09}, we automatically synthesize \indsets from programs. 
 In contrast to~\cite{Clark05qif,backesleak09,KopfR10}, the correctness of our \indsets is also automatically and machine-checked. % using refinement types.

\paragraph{Declassification}
% \paragraph{Declassification}
%
Declassification is used in IFC systems to selectively allow leaks, and several extensions of non-interference account for it~\cite{askarov2007gradual,guarnieri2019information, van2015very}; we refer the reader to~\cite{sabelfeld2009declassification} for a survey of declassification in IFC.
Most systems treat declassification statements as \emph{trusted}.
Our work focuses  on the \emph{what} dimension of declassification, that is, our policies restrict \emph{what} information can be declassified.
In contrast, \citet{chong2004security} enforce declassification policies that target other aspects of declassification, specifically, limiting in which context declassification is allowed and how data can be handled after declassification.
\paragraph{Program Synthesis}
\tool's synthesis technique follows sketch-based synthesis~\citep{solar2006combinatorial}, where 
traditionally users provide a partial implementation with \emph{holes} and some specifications based on which the synthesizer fills in the holes. 
%This is inspired by sketch-based synthesis~\citep{solar2006combinatorial}, where users provide a partial implementation containing \emph{holes} which a specification and the synthesizer  fills in the holes in a way compliant with the specification. 
%
Standard types have extensively served as a synthesis template often
combined with
tests, examples, or user-interaction~\citep{osera2015type,guria2021rbsyn,feng2017component,lubin2020program}.
Refinement types provide stronger specifications, thus, 
as demonstrated by
\textsc{Synquid}~\citep{polikarpova2016synquid}, 
do not require further tests or user information. 
%
%The intersection of
%these two ideas---partial programs and typed holes has been explored as
%well in Hazel~\citep{lubin2020program} to build a live programming
%environment. 
In \tool, we use the refinement type synthesis idea of \textsc{Synquid}, but also 
mechanically generate the knowledge specific refinement types.
% These constraints together with numerical optimization directives to Z3,
% helps to find optimal posteriors.

\section{Conclusion \& Further Applications}
\label{sec:conlusion}

We presented \tool, a novel technique that uses 
the abstractions of refinement types to 
synthesize and statically machine-check correct 
approximations of knowledge and \indsets. 
Using 
% this abstract knowledge, 
these approximations of knowledge, we defined 
a bounded downgrade function 
that can be staged on top of existing IFC systems 
to enforce declassification policies. 
We implemented \tool and demonstrated it runs across a variety of
benchmarks from prior work and can securely answer multiple sequential queries
without losing precision. We believe \tool represents a promising
approach to embedding declassification policies in applications.

%\paragraph{Policies beyond declassification}
Though we only used \tool's 
precise, explicit representation of knowledge 
for declassification, such a representation is
at the core of many information flow control tasks. Enforcing
probabilistic policies requires combining knowledge, computed by \tool,
with a probability distribution~\cite{mardziel-beliefpol-2013}. Moreover,
dynamic security policies can be enforced by keeping track of attacker
knowledge and comparing it with the current
policy~\cite{guarnieri2019information}. Finally, approximations of classical
quantitative information flow measures, such as Shannon
entropy~\cite{shannon2001mathematical}, can be derived from the
user's knowledge, \ie by counting the number of concrete elements
represented by the knowledge.

\begin{comment}
\tool opens up three lines of work we plan to investigate in the future. 
First, the knowledge-based specifications can be combined with 
state-of-the-art static security systems to encode quantitative 
policies, including declassification and selective information leak. 
Second, we plan to investigate 
how 
% if it is possible 
to encode 
knowledge as a probabilistic distribution. This encoding is common, 
but challenging to encode using deterministic refinement types. 
Finally, we plan to use relational and differential reasoning systems 
to allow formal encoding of precision and quantitative properties 
of our constructed abstract domains. 
\end{comment}

\begin{acks}
%
% \nv{Thank Hicks and DVH if gave advice?}
Thanks to Michael Hicks, David Van Horn, and Jeffrey S. Foster for their valuable advice and feedback.
We would also like to thank our shepherd Yu Feng and the anonymous reviewers for their helpful comments.
This work was supported by 
% \mg{Add all the other funding sources!}
a gift from Intel Corporation, 
NSF grants CCF-1846350, CCF-1900563, and CNS-1801545,
the Juan de la Cierva grants FJC2018-036513-I and IJC2019-041599-I, 
Madrid regional grants S2018/TCS-4339 BLOQUES and 2019-T2/TIC-13455,
Spanish national project RTI2018-102043-B-I00 SCUM, 
the HaCrypt ONR project N00014-19-1-2292, 
and the ERC Starting grant CRETE (101039196). 
\end{acks}

\balance
\bibliography{ref}
\end{document}